\pdfoutput=1

\documentclass[preprint2]{emulateapj}

\newcommand{\kms}{km s$^{-1}$}
\newcommand{\hi}{H{\sc i}}
\newcommand{\co}{$^{12}$CO(1-0)}
\newcommand{\CO}{$^{12}$CO(2-1)}
\newcommand{\halpha}{H$\alpha$}
\newcommand{\hii}{\relax \ifmmode {\mbox H\,{\scshape ii}}\else H\,{\scshape ii}
\fi}
\newcommand{\hbeta}{H$\beta$}
\newcommand{\lsol}{L$_\odot$}
\newcommand{\msol}{M$_\odot$}
\newcommand{\mi}{$\mu$m}
\newcommand{\spitzer}{{\it  Spitzer}}

\shorttitle{Molecular gas and dust in J1023+1952}
\shortauthors{Lisenfeld et al.}

\begin{document}


\title{Molecular gas and dust in Arp~94: The formation of a recycled galaxy
in an interacting system}


\author{U. Lisenfeld}
\affil{ Dept. F\'isica Te\'orica y del Cosmos, Universidad de Granada, Spain \& 
Instituto de Astrof\'\i sica de Andaluc\'\i a, CSIC, Apdo. 3004, 18080 Granada, Spain}

\author{C.G. Mundell\altaffilmark{1}}
\affil{Astrophysics Research Institute, Liverpool John Moores University, Twelve Quays House, Egerton Wharf, Birkenhead, CH41 1LD, U.K.}

\author{E. Schinnerer}
\affil{Max-Planck-Institut f\"ur Astronomie,
K\"onigstuhl 17, 69117 Heidelberg, Germany.}

\author{P. N. Appleton}
\affil{NASA Herschel Science Center (NHSC), California Institute of Technology, MC 100-22, 1200 E. California Boulevard, Pasadena, CA 91125}
\and

\author{J. Allsopp}
\affil{Astrophysics Research Institute, Liverpool John Moores University, Twelve Quays House, Egerton Wharf, Birkenhead, CH41 1LD, U.K.}

\email{ute@ugr.es, cgm@astro.livjm.ac.uk, schinner@mpia-hd.mpg.de, apple@iapc.caltech.edu}


\altaffiltext{1}{Royal Society University Research Fellow.}


\begin{abstract} 
We present  new results for the molecular gas, dust emission and the ionized gas
in J1023+1952, an \hi\ rich intergalactic star-forming tidal
dwarf galaxy candidate. It is  located at the projected intersection of
two faint stellar tidal streams wrapped around the interacting pair of
galaxies NGC~3227/6 (Arp~94).
Using the IRAM 30m telescope, emission from
\co\  and \CO\  was detected across the entire extent of
the neutral hydrogen cloud associated with J1023+1952, a region of the size of
8.9$\times$5.9~kpc,  as well as in the nuclear region and
outer disk of NGC~3227.
The molecular gas is found to be abundant over the entire \hi\ cloud, with 
H$_2$-to-\hi\ gas mass ratios between 0.5 and 1.7.
New \spitzer\ mid-infrared observations at 3.6, 4.5, 5.8, 8.0, 15 and 24 \mi\  show that
young SF  is restricted to the southern part of the cloud. 
Despite the relatively uniform H$_2$ and \hi\ column density across the cloud, young SF 
occurs only in those regions where the velocity dispersion in the CO and
\hi\ is a factor of $\sim$two lower (FWHM of 30 - 70 \kms) than elsewhere in the cloud
(FWHM of 80 - 120 \kms). Thus the
kinematics of the gas, in addition to its column density,
seems to be a crucial factor in triggering SF.
Optical/infrared spectral energy distributions (SEDs) and \halpha\ photometry
confirm that all the knots  are young, 
with a tentative age sequence from the south-west (oldest knots) to
the north-east (youngest).
Optical spectroscopy of the brightest SF region allowed us to determine
the metallicity (12+log(O/H) =  8.6 $\pm$ 0.2) and the extinction
($A_B$=2.4).
This shows that J1023+1952 is made from metal-enriched gas which
is inconsistent with the hypothesis that it represents a pre-existing dwarf galaxy. 
Instead, it must be formed from recycled, metal-enrichd gas, expelled from
NGC~3227 or NGC~3226 in a previous phase of the interaction.

\end{abstract}





\section{Introduction}

Merger-driven galaxy evolution is  important on a
wide range of scales at different cosmic epochs and in different
environments. The observed decline in the global star formation rate (SFR)
at redshifts z$<$2 and the increased dominance of small galaxies in
the local Universe has been suggested to imply interaction-driven
'downsizing' of galactic structure and star formation (Cowie et
al. 1996, Madau et al. 1996, Brinchmann et al.  1998, Treu et al.  2005, Thomas et al. 2005) 
in contrast to
hierarchical galaxy formation. The recently-discovered free-floating
intergalactic and extraplanar \hii regions around nearby gas-rich
galaxies (Gerhard et al. 2002, Ryan-Weber et al. 2004) and
H$\alpha$-emitting star-forming knots in galaxy clusters (Sakai et
al. 2002) are the smallest intergalactic structures, and may
represent an important evolutionary link or even a new mode of star formation
(SF)  in the low-density
environment of the far outskirts of the galaxy halo. Their young
star-formation ages suggest they are tidal in origin, forming from
gaseous tidal debris in intergalactic space.

The largest \hii regions are typically found at the apparent tip of 
tidal tails, at more than 100 kpc from the parent galaxies,  in maxima of the \hi\ column
density where we also detect large quantities of
molecular gas (Braine et al. 2000, 2001, Lisenfeld et al. 2002) and
where enough tidal material is often available to build a dwarf
galaxy (Tidal Dwarf Galaxies, hereafter TDGs).
%
Indeed, tidal interactions between approximately equal-mass gas-rich galaxies have long been
known to result in dramatic tidal damage, producing bridges and tails
of debris extending out to several galaxy diameters (e.g. Toomre \&
Toomre 1972, Haynes et al. 1984, Barnes \& Hernquist 1992, Mihos
2001, Mundell et al. 1995).  Tidal tails often comprise primarily
neutral hydrogen since \hi\ is usually the most spatially extended
 component of a spiral galaxy disk, and is therefore most
 easily stripped during the earliest stages of a tidal encounter
(e.g. Hibbard \& van Gorkom 1996). 
Molecular gas is generally  observed to be more centrally concentrated than \hi\ in disks of
galaxies outside clusters (Nishiyama, Nakai \& Kuno 2001)
so it is less likley to be stripped on first passage.

Here we present a study of an extragalactic region with active SF  embedded in
an extensive gas cloud named J1023+1952.
The region is associated with the interacting Seyfert
system Arp~94 (Mundell et al. 1995; Mundell et al. 2004). Optically,
the system is dominated by two disturbed Seyfert galaxies: NGC~3227,
an SAB(s) pec barred spiral Seyfert galaxy and its elliptical
companion, NGC3226 (E2 pec) (Rubin \& Ford 1968, Mundell et
al. 2004).  Neutral hydrogen (\hi) imaging of the system (Mundell et
al. 1995) revealed two gaseous tidal tails extending $\sim$100~kpc
north and south of NGC~3227, well-ordered gas in the disk of NGC~3227
and a massive \hi\ cloud (M$_{\rm
\hi}$~=~3.8~$\times$~10$^8$~M$_{\odot}$), that lies at the base of the
northern tail and is close to, but physically and kinematically
distinct from the disk of NGC~3227. Mundell et al. (1995) suggested
that this cloud (hereafter J1023+1952) might be a dwarf galaxy that is
either pre-existing and being accreted by Arp~94, or a newly-created
tidal dwarf galaxy (TDG) forming from the tidal debris.  

The subsequent discovery of a region of very blue star-forming knots
with luminosities between $10^6$ and $5\times 10^6$ \lsol, 
embedded in a high \hi\ column density ridge in the southern half of
J1023+1952 (Mundell et al. 2004) confirmed its classification as a
dwarf galaxy; the inferred SFR, UV luminosities and
H$\alpha$ equivalent widths of the knots suggested a starburst age
less than 10~Myr. Near infrared imaging provided further evidence for
the youth of J1023+1952; no additional embedded star formation or old
stellar population were found (Mundell et al. 2004). 
These results suggest that the observed SF is currently being triggered in the 
interstellar medium (ISM) of J1023+1952 and does not merely represent young 
stars tidally stripped from the disk of NGC3227 (Mundell et al. 2004).


The Arp 94 interacting system offers an interesting nearby laboratory in which to study the interplay between SF and the physics of the multiphase ISM in a tidally disrupted system with young, active intergalactic SF regions. Here we present a study of 
the optical, infrared and millimetre properties of the Arp 94 system that aims to determine the nature and origin of J1023+1952, its role in the observed interaction and the underlying reasons for its observed star formation properties.
Our dataset comprises (a) \co\ and \CO\ mapping of the full extent of J1023+1923, the nucleus of NGC3227 and the outer edges of the galactic disk of NGC3227; (b) multi-band infrared imaging with the {\it Spitzer Space Telescope}  of J1023+1952 with particular emphasis on investigating the known SF regions and searching for further embedded SF; (c) optical spectroscopy of the brightest \hii region to determine the gas-phase metallicity.

The paper is structured as follows:
the observations and data reduction are described in Section 2, molecular and atomic gas and star formation properties are derived in Section 3 and the implications for the
triggering of SF and  the origin of J1023+1952 as well as its role in the system are discussed in Section 4. 
We conclude in Section 5.
Assuming a distance of 20.4 Mpc to the Arp~94 system (Tully 1988),
1\arcsec corresponds to  $\sim$100~pc.

\section{Observations and analysis}

\subsection{CO data}

We observed the \co\ and \CO\  lines at 115 and 230\,GHz
in June 2004 with the IRAM 30-meter telescope on Pico Veleta.  
We fully mapped the \hi\  extent of J1023+1952 and obtained
two spectra towards NGC~3227 (see Figure \ref{spectra} (left)
for the positions of the individual pointing).
%

Dual
polarization receivers were used at both frequencies with the 512
$\times$ 1 MHz filterbanks on the CO(1--0) line and the
256 $\times$ 4 MHz filterbanks on the CO(2--1).
Relatively high water-vapor led to a high system temperature for the 230 GHz data, and this degraded the signal-to-noise and baseline quality of the data. As a result,  we could not further
analyse the individual spectra, instead we averaged the emission in the CO(2--1) line over the whole cloud at this wavelength.
The observations were done in wobbler switching mode with a
wobbler throw of 200\arcsec \ in azimuthal direction.
At the
beginning of the observations  the frequency tuning was checked by
observing the central position of NGC~3227.
Pointing was
monitored on nearby quasars or Jupiter every 60 -- 90  minutes.
During the observation period, the weather conditions were generally good (with pointing 
better than 3~\arcsec), but we sometimes ($\sim$~4\% of the total time) experienced periods of
 anomalous refractive index resulting in large ($>$ 5\arcsec) pointing uncertainties. 
 These data were excluded from our analysis.
The average system temperature
was 390 K at 115\,GHz
on the $T_{\rm A}^*$ scale.
At 115 GHz (230 GHz),  the IRAM forward
efficiency, $F_{\rm eff}$, was 0.95 (0.91), the 
beam efficiency, $B_{\rm eff}$, was 0.75 (0.54), and the
half-power beam size  22$^{\prime\prime}$ (11\arcsec).
All CO spectra and luminosities are
presented on the main beam temperature scale ($T_{\rm mb}$) which is
defined as $T_{\rm mb} = (F_{\rm eff}/B_{\rm eff})\times T_{\rm A}^*$.
For the data reduction, we selected the observations taken
during satisfactory weather condition (no anomalous refraction, low
opacity), summed the spectra over the individual
positions and subtracted a constant continuum level.

Although the edge of the galactic disk of NGC~3227 and J1023+1952
spatially overlap in projection (Mundell et al. 1995; Mundell et
al. 2004), it is possible to disentangle their different kinematic
components in the current \co\ dataset. On average, the
emission from NGC~3227 is blue-shifted by approximately 200 \kms\
with respect to that of J1023+1952, with linewidths not in excess of
100 km s$^{-1}$, making it possible to separate them by fitting a
double Gaussian line profile. The spectra were Hanning
smoothed before the Gaussians were fitted. Initial fitting parameters
were estimated by searching for maximum values and their associated
positions in the spectra in two 150 km s$^{-1}$ wide regions each centred
on -200 and 0.0 km s$^{-1}$ with respect to the systemic velocity
of J1023+1952 of 1260 \kms. Based on examination of the spectra by eye,
the initial FWHM of the lines were set to 180 and 120 km s$^{-1}$ for
NGC~3227 and J1023+1952 respectively.

As the two objects do not spatially overlap over the entire region mapped, three
types of fit were undertaken, a simultaneous double fit, and 
two separate fits, each centred on the systemic velocity of 
one  object. Chi-squared values for each of the fits at each
spatial pixel were calculated and were filtered according to the following
criteria: (i) a line must be in emission (i.e. negative components are
rejected as unphysical), (ii) the line centres must lie in a window
from -300 to 300 km~s$^{-1}$, (iii) the FWHM must be greater than 18
km s$^{-1}$ but less than 490 km s$^{-1}$, (iv) the fit must be statistically
significant at the 5-$\sigma$ level or above. The significance limit was
calculated from the noise measured at emission-free regions of the
spectra and divided by the square root of the number of channels
across the FWHM of the line. The median noise in the data was found to
be 9.6~mK per 2.6 \kms \ channel
with mean values approaching zero, indicating good
baselining. The fit with the lowest chi-squared value meeting these criteria
was then selected.
The resulting fits for the individual components  are presented in Fig.~\ref{fits},
showing that satisfactory fits to the data could be achieved.
The resulting difference between fit and data (lower part of each panel) 
was within the noise and showed a constant level over the whole
velocity range.
Further support for the correct decomposition of the CO lines comes
from the good agreement between CO and \hi\ line velocities and widths
which will be discussed in Sect. 3.

\subsection{Optical spectroscopy}

The optical spectra were obtained on May 9, 2007 with the
instrument CAFOS at the 2.2m telescope at 
Calar Alto. The slit had a fixed width of 1.2\arcsec\  and was placed
with a position angle of 82$^\circ$, crossing \hii regions 1 and 1a (the 
labelling of the regions will be defined in Sect. 3.3.1) and
the nuclear region of NGC~3227. The spectra at knot 1a were too
weak to derive the extinction or metallicity and are not taken into account.
In order to cover the entire
wavelength range, 2 spectra with different grisms, B100 (covering the range between
3200 -- 5800 \AA, with a dispersion of 2 \AA/pixel)
and G100 (covering the range between
4900 -- 7800 \AA, with a dispersion of 2.12 \AA/pixel) were taken. 
The total integration time in each grism was 3600s and 3400s, respectively.
The data reduction was carried out in a standard way using the IRAF software.
The spectra were flux-calibrated using spectrophotometric standard
stars, under non-photometric, but transparent conditions suitable for relative
line-flux measurements. An example of the final ``red" spectrum is
shown in Fig.~\ref{opt-spec}.

\subsection{Spitzer data}

The \spitzer\ observations presented in this paper encompass
observations covering both NGC 3226 and NGC 3227 which will be 
presented in Appleton et al. (in prep.). 
They were made early in the Spitzer mission with all three instruments, IRAC (2003-Nov-26), MIPS (2003-Nov-24) and IRS-Peakup (2003-Nov-29).
 IRAC imaging in all four bands (3.6, 4.5, 5.8 and 8
$\mu$m) was performed using a 7-point dither pattern to encompass
both NGC~3227 and NGC~3226. Similarly, using a small-map photometry mode,
images were made with MIPS at 24, 70 and 160 $\mu$m, however only
the images at 24$\mu$m were useful for J1023+1952 and are included in this
study.
Imaging was also performed using the IRS blue and red peakup apertures
(IRS-PUI). 
IRAC and MIPS maps were made using the S14 (IRAC) and S15 (MIPS) data pipelines at the 
Spitzer Science Center. The MIPS results were checked by performing off-line mosaicing with different methods of background subtraction, but the results were similar to those obtained in the pipeline. An exception was the IRS-peak-up imaging in which the final image was created by running Basic Calibrated Data frames (BCDs) created in the S14 IRS pipeline offline using the SSC MOPEX software. This produced good results for the background in the region of J1023+1952.

At 24 $\mu$m the central AGN of NGC~3227 was so strong that 
its point-spread function (PSF) extended out to the region of J1023+1952,
visible in symmetrical radial fingers.
We corrected for this by subtracting a point source convolved with the PSF
of MIPS at the position of 
the central AGN. 
The removal of the PSF of
the AGN at 24$\mu$m resulted in a roughly constant decrease in flux from the
direction of J1023+1952 of approximately 30\%, and this is taken into
account in the uncertainties quoted.
In Fig.~\ref{spitzer-24} the Arp 94 image at 24 $\mu$m before and
after the AGN substraction is shown. 
At 16$\mu$m the PSF of the bright AGN was also visible but did not
extend into the area of the \hi\ cloud so that the original image could
be used.
%
%


The 8 $\mu$m emission contains emission from  stellar continuum
as well as dust emission, mainly from PAH emission.
We estimate the contribution from dust by subtracting the
stellar emission, applying the formulae
given in P\'erez-Gonz\'alez et al.  (2006). They  
followed the procedure described in Pahre et al. (2004)
who
used a Vega star template to extrapolate the
stellar light from 3.5/4.5 to 8 $\mu$m. 
We found that in all individual knots the 8$\mu$m 
emission was almost entirely  ($>$ 90\%) due to dust. Therefore,
we neglect in the following a possible small
contamination from stellar light and assume that
the entire 8~$\mu$m emission is from dust.

\section{Results}

\subsection{Molecular and atomic gas}

\subsubsection{Gas distribution}

\begin{figure*}[h!]
\epsscale{1.0}
\plotone{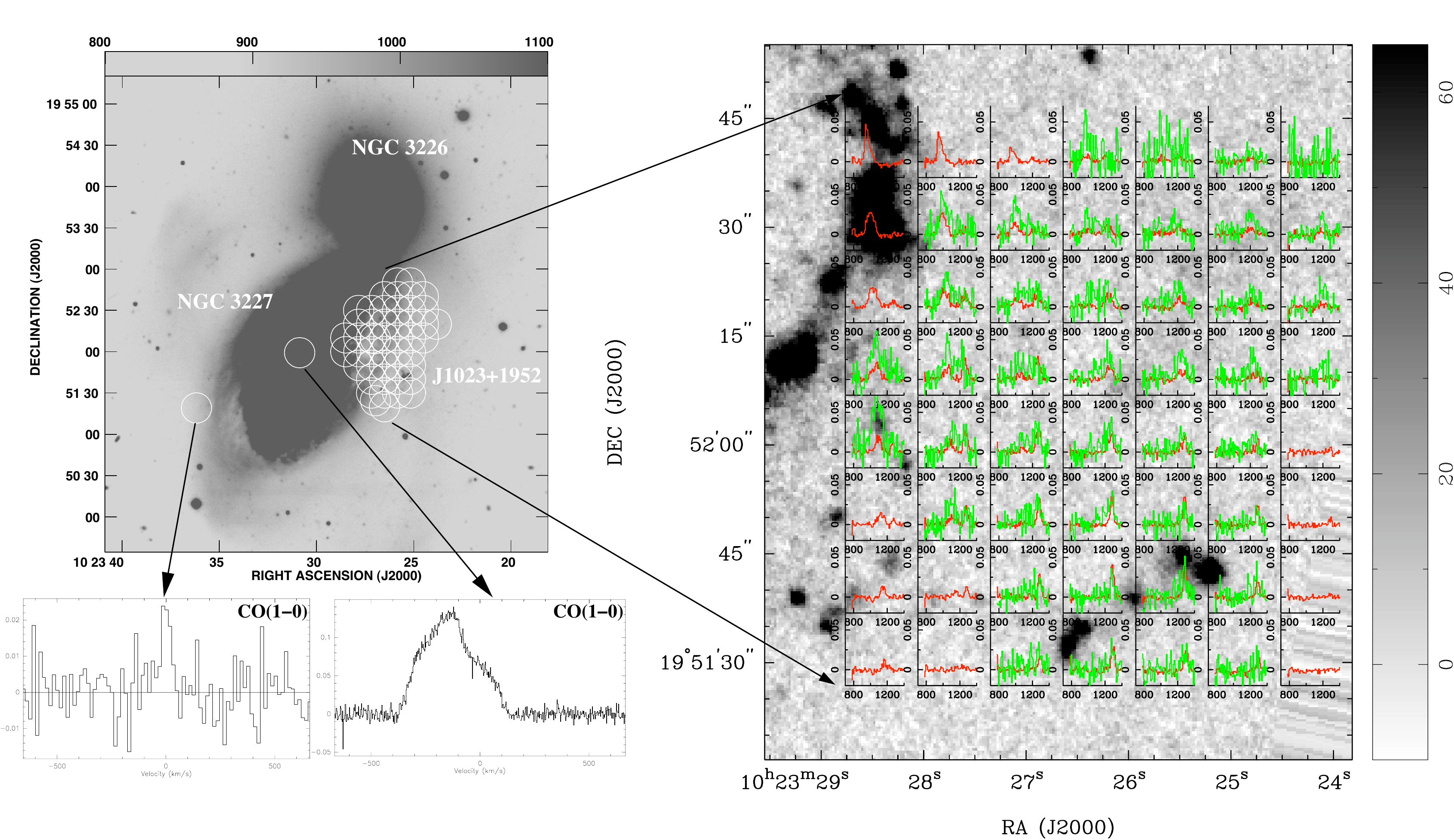}
\caption{{\it Left:}
Position of the IRAM pointings, overlaid
on a blue image, together with the CO spectra of the
positions observed within the disk of NGC 3227.
{\it Right:} Spectra of \hi\ emission (red) and $^{12}$CO(1-0) 
emission (green) 
from J1023+1952 overlaid on an H$\alpha$ image from Mundell et al. (2004).
\label{spectra}}
\end{figure*}

\begin{figure*}[h!]
\epsscale{1.0}
\plotone{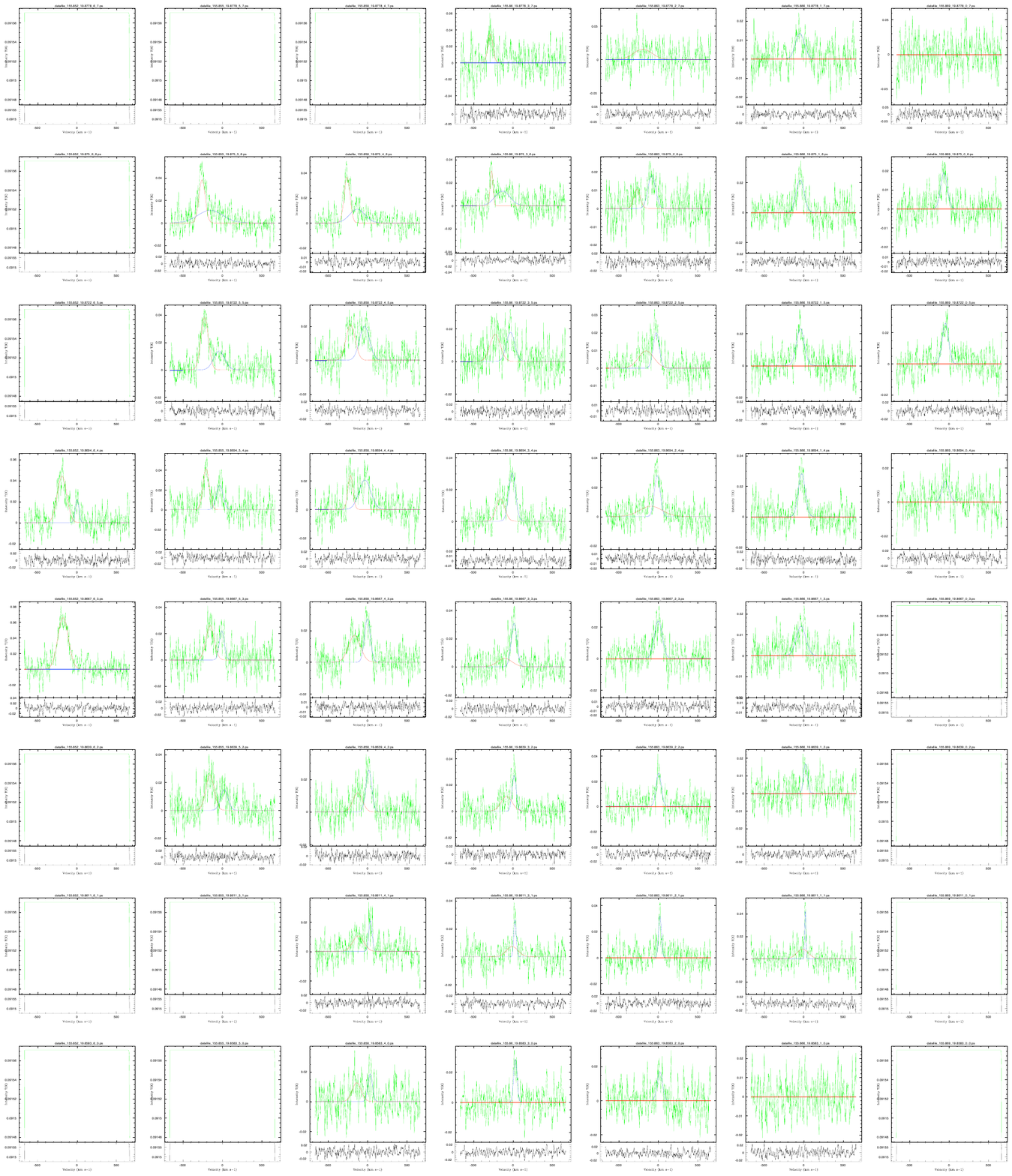}
\caption{Observed spectra (green) together with their Gaussian fits. The red line shows the fit
to the component from the disk of NGC 3227 and the blue line to the 
component of J1023+1952. The velocity scale
is relative to a systemic velocity of 1260 \kms. The individual
panels correspond to the observed positions shown in Fig.\ref{spectra}.
The lower plot in each panel shows the difference between the 
two-component fit and the data.
Note the narrower linewidths coincident with the star
forming knots in J1023+1952 (lowest two rows of spectra, see Fig. 1).
\label{fits}}
\end{figure*}

\begin{figure}[h!]
\epsscale{1.0}
\plotone{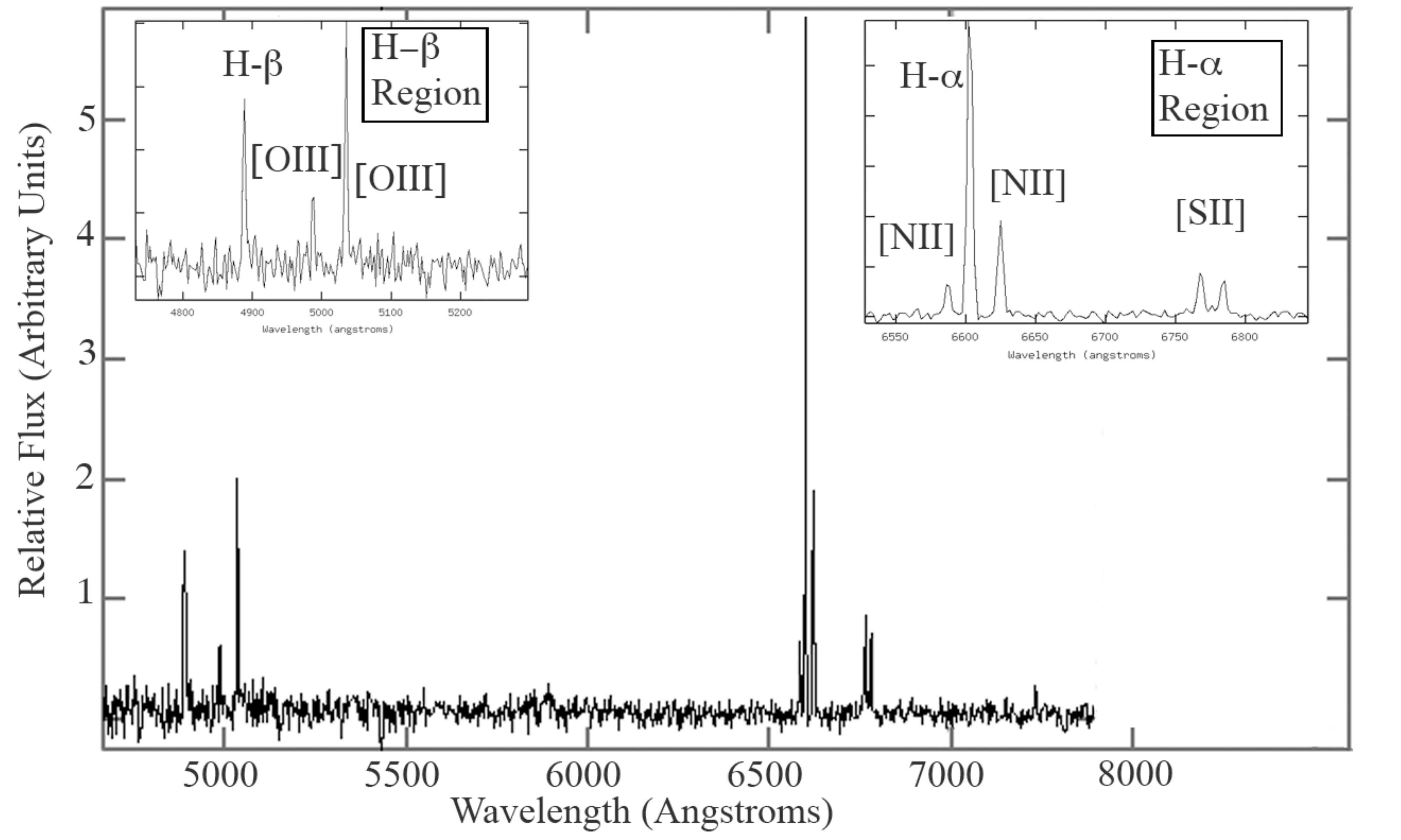}
\caption{The optical spectrum at the position of the knot 1. The insets show
blown-up versions of the relevant regions around the \halpha\ and H$\beta$ line.
\label{opt-spec}}
\end{figure}

\begin{figure}[h!]
\epsscale{0.5}
\centerline{\plotone{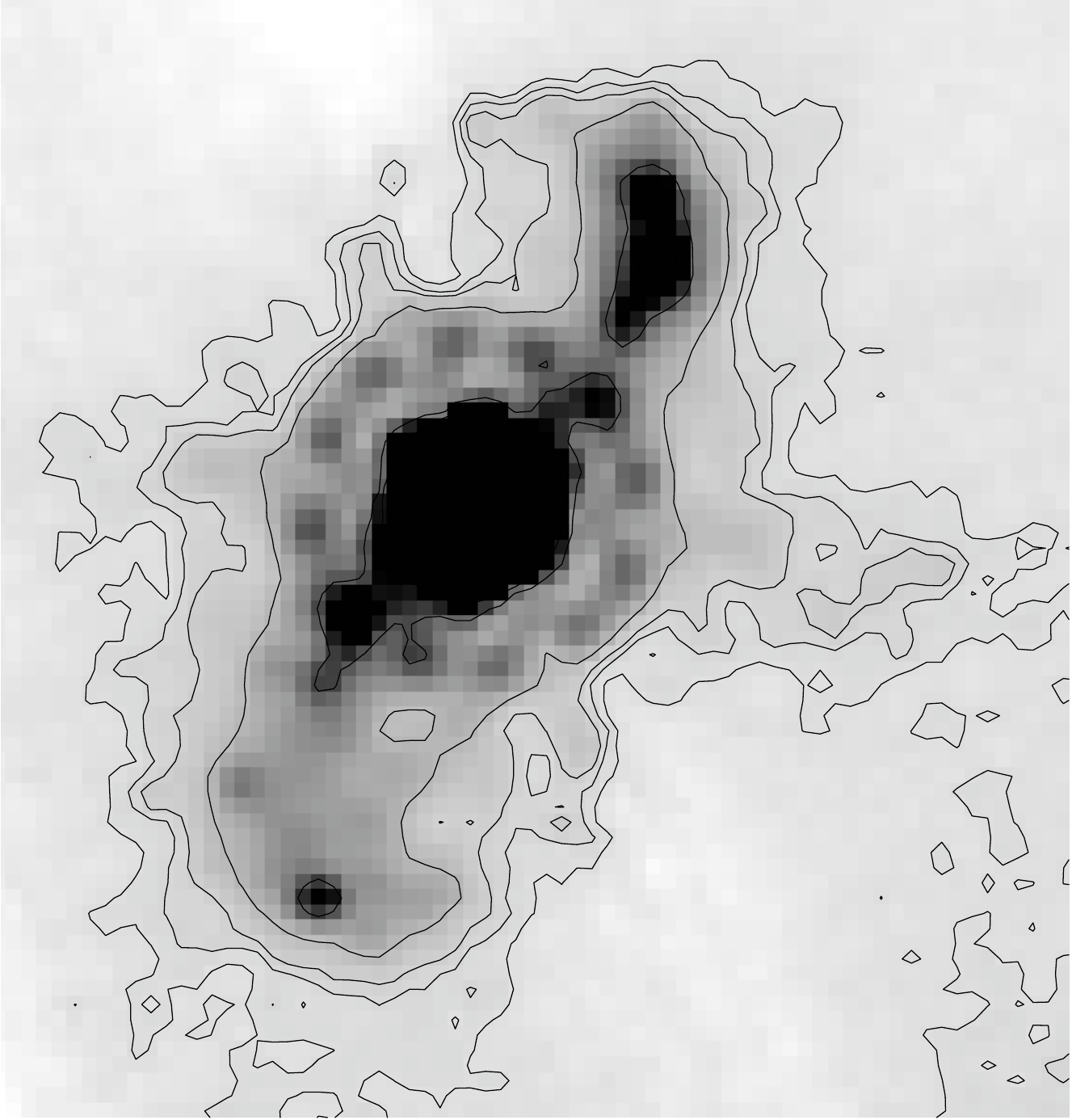}
\plotone{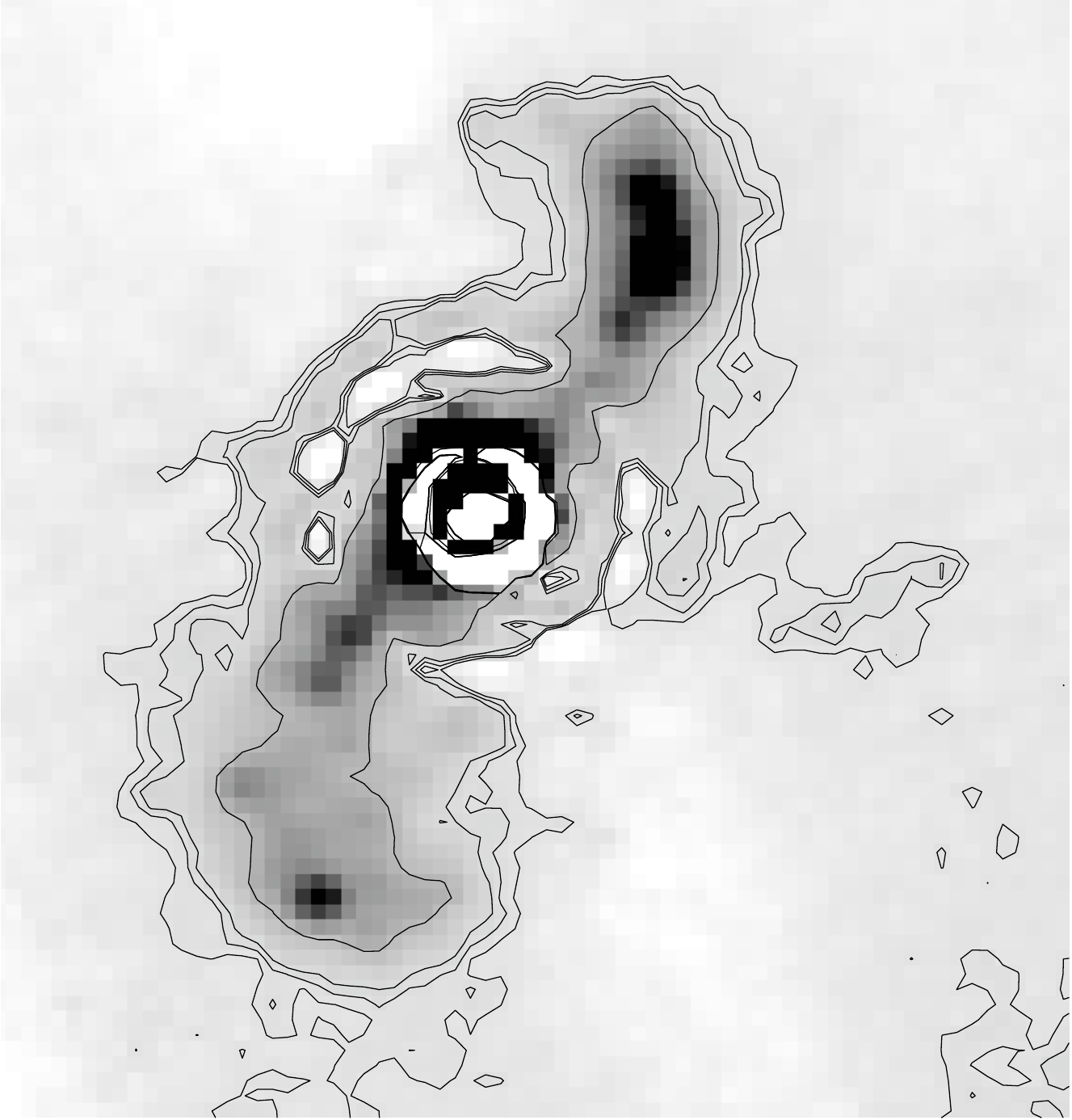}}
\caption{{\it Left:} 24\mi\ image of NGC 3227 and J1023+1952.
{\it Right:} Sidelobe emission from the strong nucleus of NGC 3227
has been subtracted and the emission at the SF region of J1023+1952 is now clearly visible.
\label{spitzer-24}}
\end{figure}

Figure \ref{spectra} (right) shows \co\ spectra (green) taken across the Arp~94
system, superimposed
on an \halpha\ image in which the  disk of NGC~3227 and the
star forming knots in J1023+1952 can be seen. \co\ emission is
detected (a) across the entire $\sim$9$\times$6~kpc \hi\ 
extent of J1023+1952, not
only in the star-forming regions, (b) from the well-studied gas-rich
Seyfert nucleus of NGC~3227 as expected (Schinnerer et al. 2000) and (c) at
the outer edges of the disk of NGC~3227 at a radius of $\sim$9~kpc to
the SE and NW of the nucleus.
Also shown for comparison in Fig. \ref{spectra} (right) are the corresponding \hi\ 
spectra (red) at these locations. 

\begin{figure}[h!]
\epsscale{1.0}
\plotone{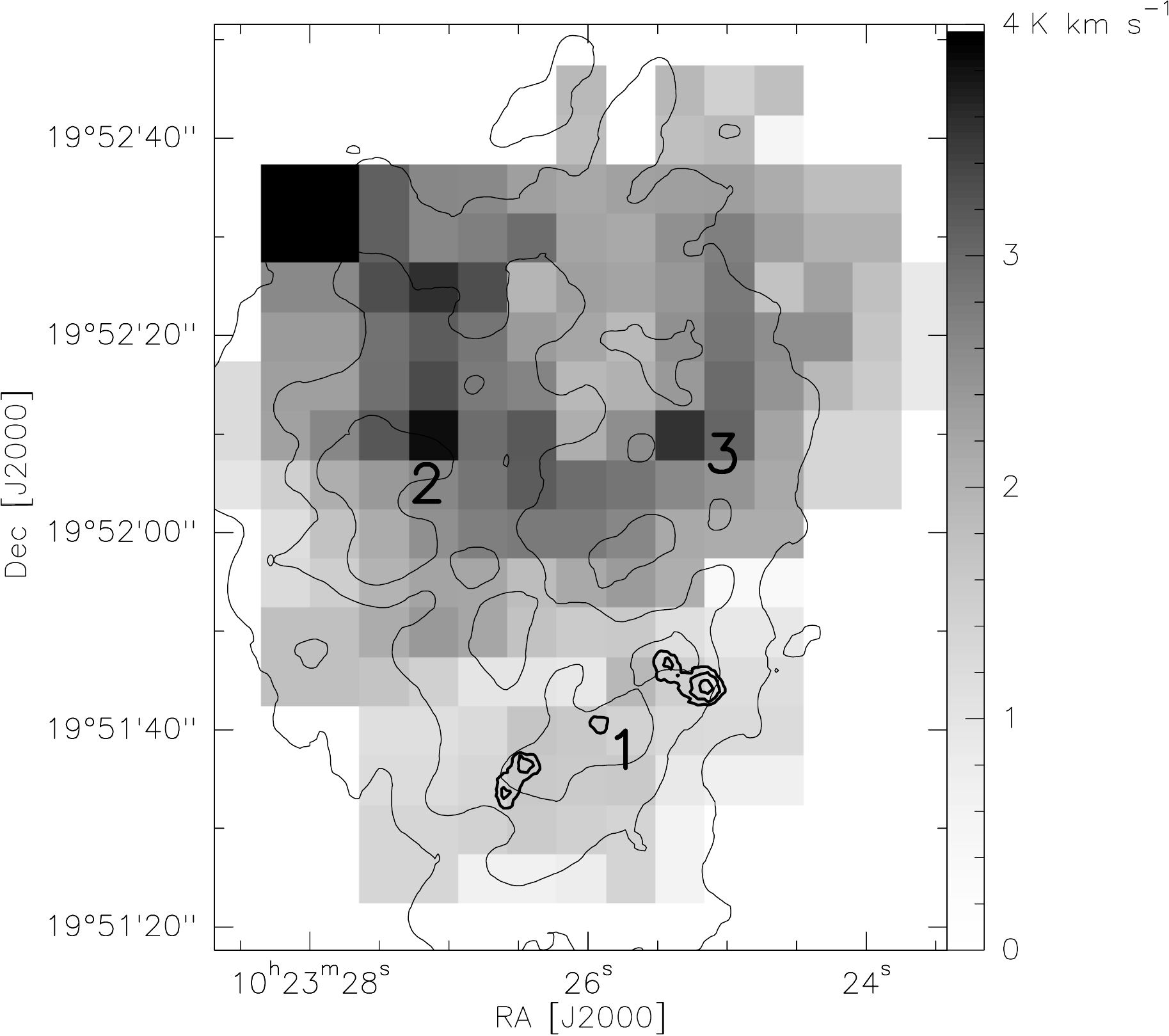}
\caption{Velocity integrated intensities (moment 0) of CO (grey); 1\, K\,\kms\ correspond to a surface density of 2$\times 10^{20}$ cm$^{-2}$.
Overlaid (contour) is the  moment 0 map of \hi\ 
(Mundell et al. 2004),  observed with the VLA in
B-array with a spatial resolution of 6.3\arcsec $\times$ 6.3\arcsec.
The contour levels are 10, 40, and 70 Jy beam$^{-1}$ ms$^{-1}$,
corresponding to 0.3, 1.1, and 2.0 $\times 10^{21}$ cm$^{-2}$.
The numbers indicate positions discussed seperately in the text.
The thick contours show the  \halpha\ emission.
\label{mom0}}
\end{figure}

Figure~\ref{mom0} shows the velocity integrated intensity of the CO emission
(which is proportional to the molecular gas column density)
of J1023+1952 alone (after separating out the emission of NGC 3227 based
on the Gaussian fit, as explained in Sect. 2.1), overlaid with contours of the \hi\ surface density
at a spatial  resolution of 6.3\arcsec$\times$ 6.3\arcsec.
It is interesting that the molecular gas surface density in the northern part is
even higher than in the southern part.
The distributions of \hi\ and H$_2$ show differences: the \hi\
peaks in a ridge-like structure towards the east which 
is not followed by the  H$_2$ which presents instead a decrease in 
surface density from north to south. In order to quantify these trends, we list
in Table \ref{surface_density} the column  densities of the 
molecular gas (assuming a standard
Galactic conversion factor of
$N(H_2)/I_{CO}=2\times10^{20}$~cm$^{-2}$~(K~\kms)$^{-1}$),
atomic gas and total  gas, as well as the ratio between the H$_2$ and \hi\
column densities (molecular gas fraction)
for the 3 regions labelled in Fig. \ref{mom0}. 
These regions correspond to: (1) the southern SF region,
(2) a region where both the atomic and molecular gas surface
density are high and (3) a region with
low atomic, but high molecular surface density.
To derive these values we used the \hi\ C-array data from Mundell et al. (1995)
which had a spatial resolution of 20.47\arcsec $\times$ 18.25\arcsec, comparable to that of the
CO data.

Surprisingly, the lowest molecular surface density and molecular
fraction are found in region 1, the SF region. This shows that 
a high molecular (or total) gas column 
density is not enough to predict whether SF takes place or not.
The \hi\ column density, on the other hand, 
has a peak in region 1. 

%

\subsubsection{CO line ratio}

\CO\ emission was also detected from J1023+1952 at most locations,
although with a poorer S/N than 
\co , but sufficient to calculate an mean ratio \CO/\co , averaged over
the whole object. 
Since the S/N of the \CO\ data was too low to perform
the separation between the emission of  J1023+1952 and NGC~3227 we 
made the approximation that emission below -120 \kms\ (for both
\co\ and \CO) was from NGC~3227 and
emission above this value from J1023+1952. In this way, we derived
an average ratio \CO/\co\ $= 0.54\pm0.10$, where the error takes into account both the error due 
to the rms noise of the spectra and 
a calibration uncertainty of 10\%. 
We do not take into account a possible error due to the undersampling
of the \CO\ emission with our 10\arcsec\ spacing.
The line ratio is practically unchanged when including the emission of NGC~3227
which shows that there is no significant difference between the line ratios
of J1023+1952 and the outer disk of NGC~3227.
This ratio is lower  than the mean value  found in the central regions of
 spiral galaxies (e.g., $0.89\pm 0.06$, with a dispersion of 0.34, Braine \& Combes 1992).
 The low line ratio could indicate a relatively low gas density, leading to a subthermal
 excitation of the \CO\ line.
Due to the different beams of the \CO\ and \co\
spectra and the undersampling of the \CO\ emission, we were not able to
study local variations of the line ratio.

\subsubsection{Gas kinematics}

\begin{figure}[h!]
\epsscale{1.0}
\plotone{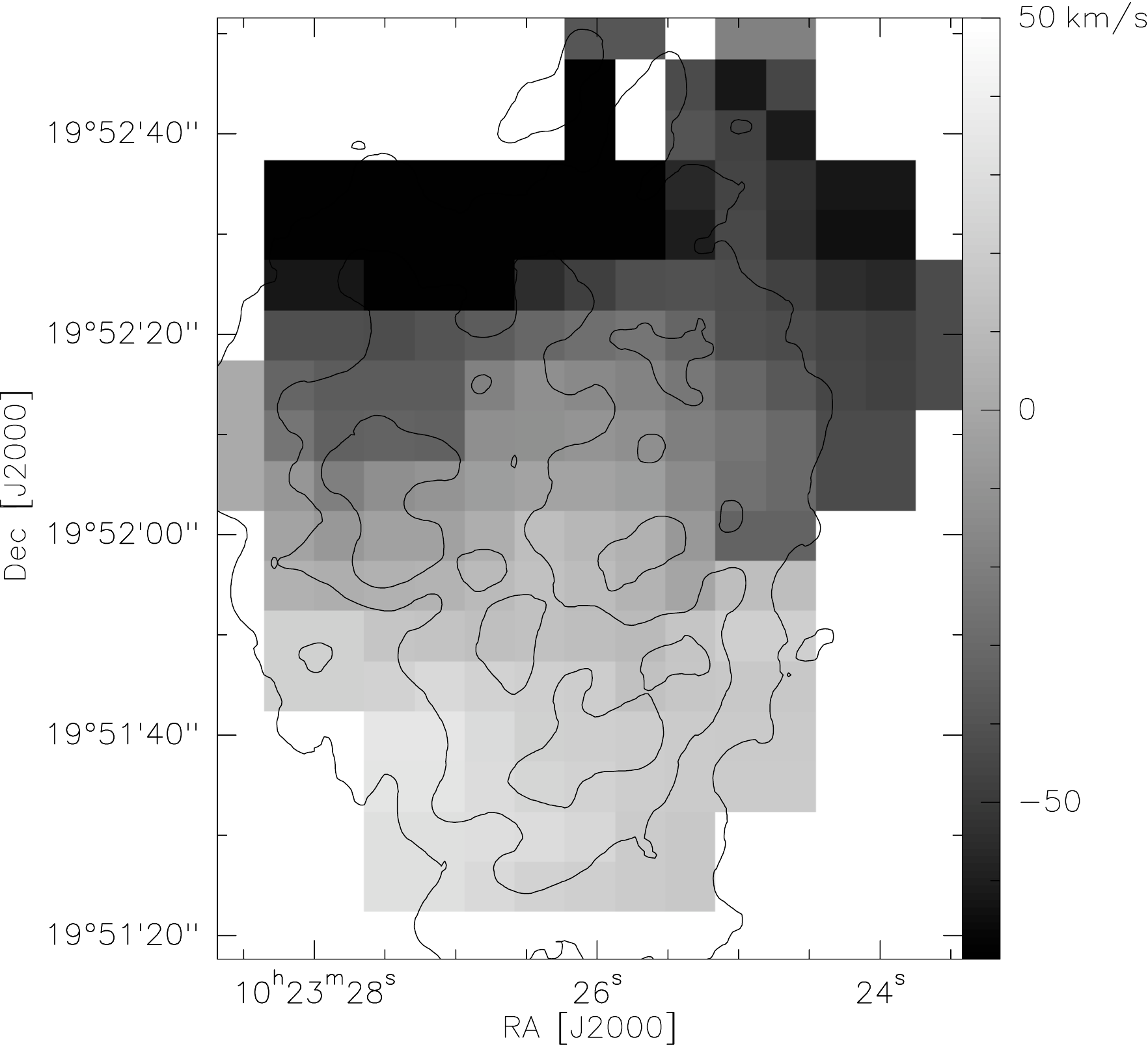}
\caption{Intensity weighted radial velocity (moment 1) of CO (grey) overlaid on
 the same moment 0 maps of \hi\ 
(contours)
as in Fig.~\ref{mom0}. The values of the intensity-weighted radial velocities
are relative to a systemic velocity of 1260 \kms .
\label{mom1}}
\end{figure}

In Fig.~\ref{mom1} we show the intensity-weigthed  radial velocities.
These velocities show a smooth 
gradient over J1023+1952 with a maximum shift of about 80 \kms\ from the north to the
south over a distance of about 70\arcsec\ (6.9~kpc).
The gradient agrees very well with that seen in the \hi\ line (see Fig. 8b in
Mundell et al. 1995).
The agreement in the velocity fields derived for the CO and \hi\ provides further
confidence that the profile decomposition between the CO emission from J1023+1952 and NGC 3227 in the north-east of the cloud was successful.

\begin{figure}[h!]
\epsscale{1.0}
\plotone{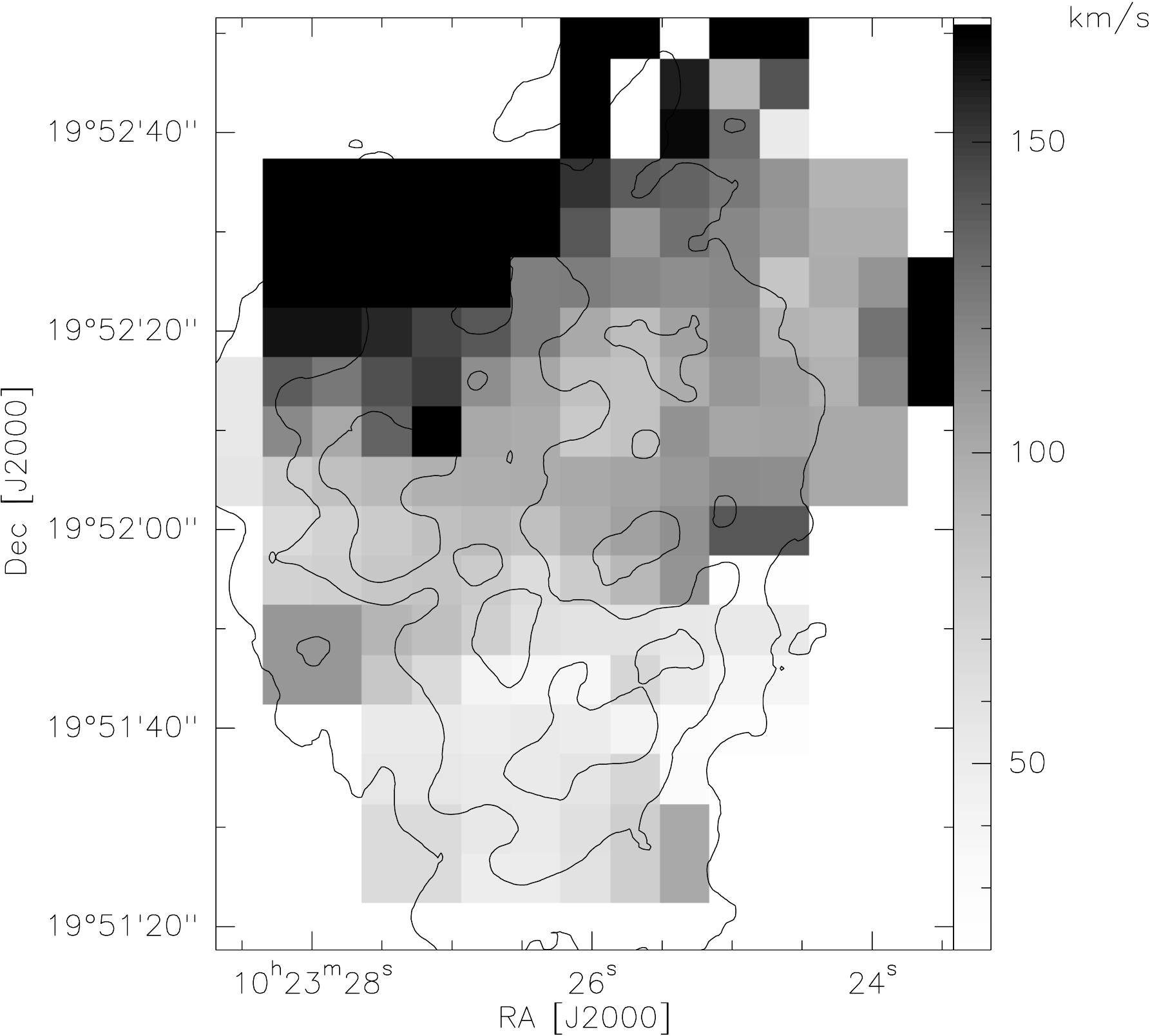}
\caption{FWHM of the CO lines (grey) overlaid on the same 
moment 0 maps of \hi\ 
(contours)
as in Fig.~\ref{mom0}.
\label{mom2}}
\end{figure}

Fig.~\ref{mom2} shows the 
line velocity widths (FWHM) of the   \co\  emission. The linewidths are smallest in
the southern SF region (region 1)  with values for the
FWZI around 100 \kms\ and FWHM of the Gaussian fit between 30 and
70 \kms. Northwards of this area (region 3),
the linewidths become substantially larger,
with FWZI around 200 \kms\ and FWHM of the Gaussian fit between
 80 and 120 \kms . Also the area around the \hi\ peak (region 2) shows
broad lines, with FWHM of 80-90 \kms .
The same trends hold for the \hi\ emission, since the \hi\ and CO spectra
agree very well in velocity and width (see Fig.~\ref{spectra}).

\subsubsection{Gas Masses}

Assuming a standard
Galactic conversion factor of
$N(H_2)/I_{CO}=2\times10^{20}$~cm$^{-2}$~(K~\kms)$^{-1}$, the total
molecular gas mass can be calculated via
$M_{H_2}[M_{\odot}]=75\,I_{CO}D^2\Omega$ (Lisenfeld et al. 2002), where $I_{CO}$ is
the velocity-integrated CO line intensity in K~\kms, D is distance in
Mpc and $\Omega$ is the area in arcsec$^2$. The integrated intensity
for J1023+1952, over the extent of the CO emission of $\sim 90$\arcsec $\times 60$\arcsec
(corresponding to 8.9$\times$5.9~kpc)
is 1.3~K~\kms, resulting in a total molecular gas mass
$M_{H_2}~\sim 2.2\times 10^{8} M_{\odot}$ (including a Helium fraction of
1.38, M$_{H_2+He}~\sim 3.0\times 10^{8} M_{\odot}$).
The total mass of the atomic gas is M$_{HI} = 3.8\times 10^{8} M_{\odot}$
(Mundell et al. 2004), giving a rather high molecular-to-atomic gas mass fraction of
$M_{H_2}/M_{HI} = 0.6$.

Including only narrow-line emission from
the vicinity of the star forming knots, results in
$M_{H_2+He}\sim 1.05\times 10^{8}$ M$_{\odot}$.
The SFR in this region, derived from the  \halpha\ emission
(extinction corrected with A$_{\rm H\alpha} = 1.4$ mag, see next Section)
is $3.6 \times  10^{-2}$ \msol yr$^{-1}$, which yields 
a gas consumption time of $3 \times  10^{9}$ yr.
Taking also atomic gas into account, the gas consumption time
increases by a factor 2.3.
This long time scale indicates that plenty of gas is available
to drive SF for an extended period of time. 

%
%

\subsection{Extinction and metallicity}

Line fluxes 
were measured in the optical spectrum shown in Fig.~\ref{opt-spec}
 using the IRAF {\it splot}
procedure. 
We used the \halpha/\hbeta\ ratio to derive the extinction
assuming case B recombination of Osterbrok (1989), which predicts an intrinsic
 \halpha/\hbeta\ of 2.86 for an electron temperature of 10\,000 K.
We derived an extinction of $A_{H\alpha} = 1.4$~mag.
Based on this value and the extinction curve given in Draine (2003) (his
Table~4), we dereddened the lines.
The resulting, dereddend line fluxes, relative to \hbeta , are 
listed in Table~\ref{optical_lines}.

We derived the oxygen abundance based on the standard empirical
R23 method (Pagel et al. 1979). From 
$\rm log([OIII]_{\lambda 5007}/[NII]_{\lambda 6584}) = 0.2$
we determined (e.g. from Edmunds \& Pagels 1984) that we are in
the upper branch of the R23 -- 12+log(O/H) relation.
The value of R23 = $\rm ([OIII]_{\lambda 5007}+[OIII]_{\lambda 4959}+
[OII]_{\lambda 3737})/H\beta) = 6.2$ yields
 then  12+log(O/H) = 8.6$\pm$ 0.2.
We also applied the empirical method of Denicol\'o et al. (2002) which is 
based on the nitrogen-to-\halpha\ ratio and derived from
$\rm log([NII]_{\lambda  6584}/H\alpha) = -0.5$ practically
the same oxygen abundance,  12+log(O/H) = 8.7$\pm$ 0.2.
This metallicity is higher than what is typically found in dwarf galaxies,
and  more similar to  values measured in disks of spiral galaxies (e.g
the solar metallicity  is 12+log(O/H) = 8.66,
Asplund et al. 2005).

\subsection{Star Formation in J1023+1952}

\subsubsection{\spitzer\ Mid-IR Emission}

\begin{figure*}[h!]
\epsscale{0.3}
\centerline{\plotone{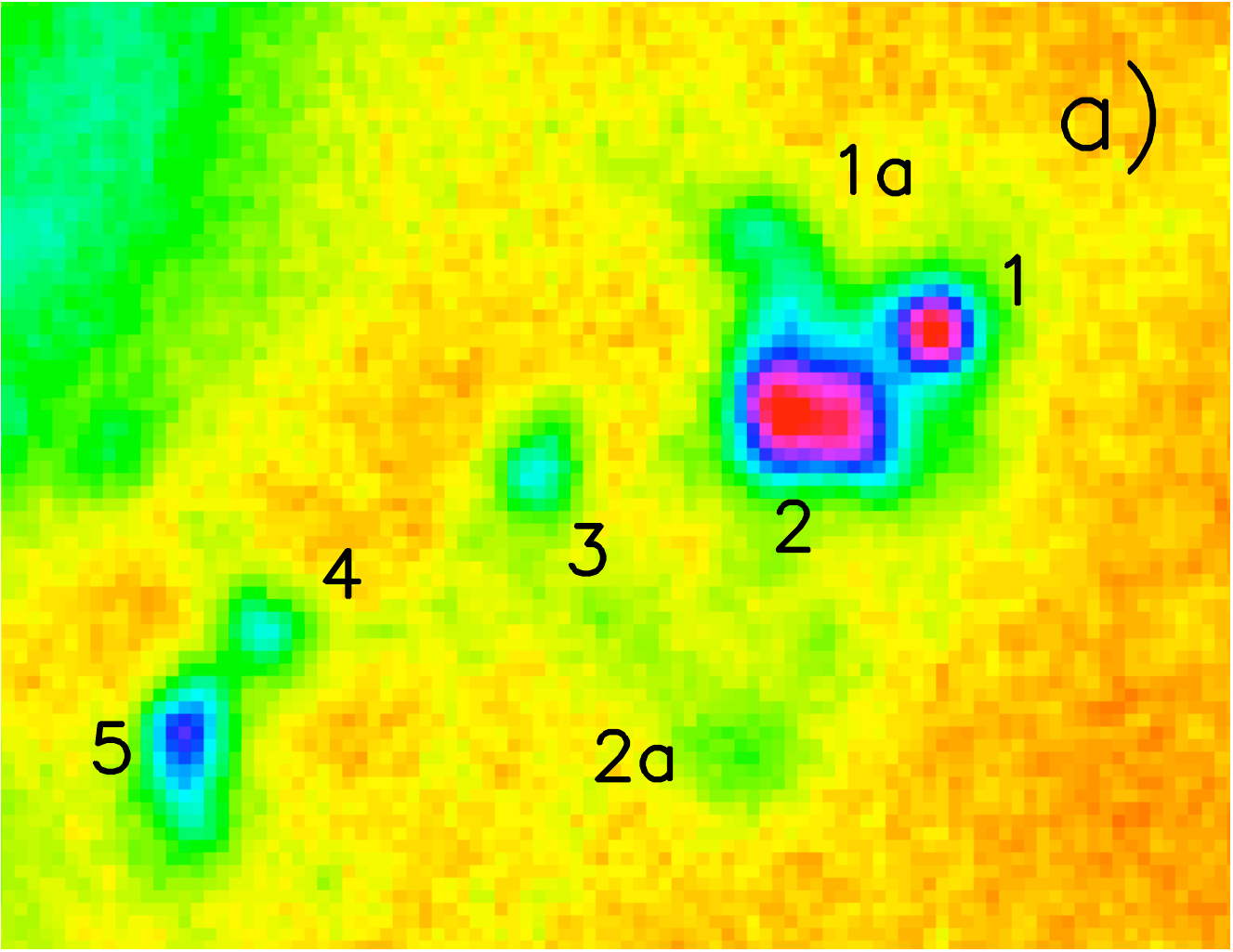}
\plotone{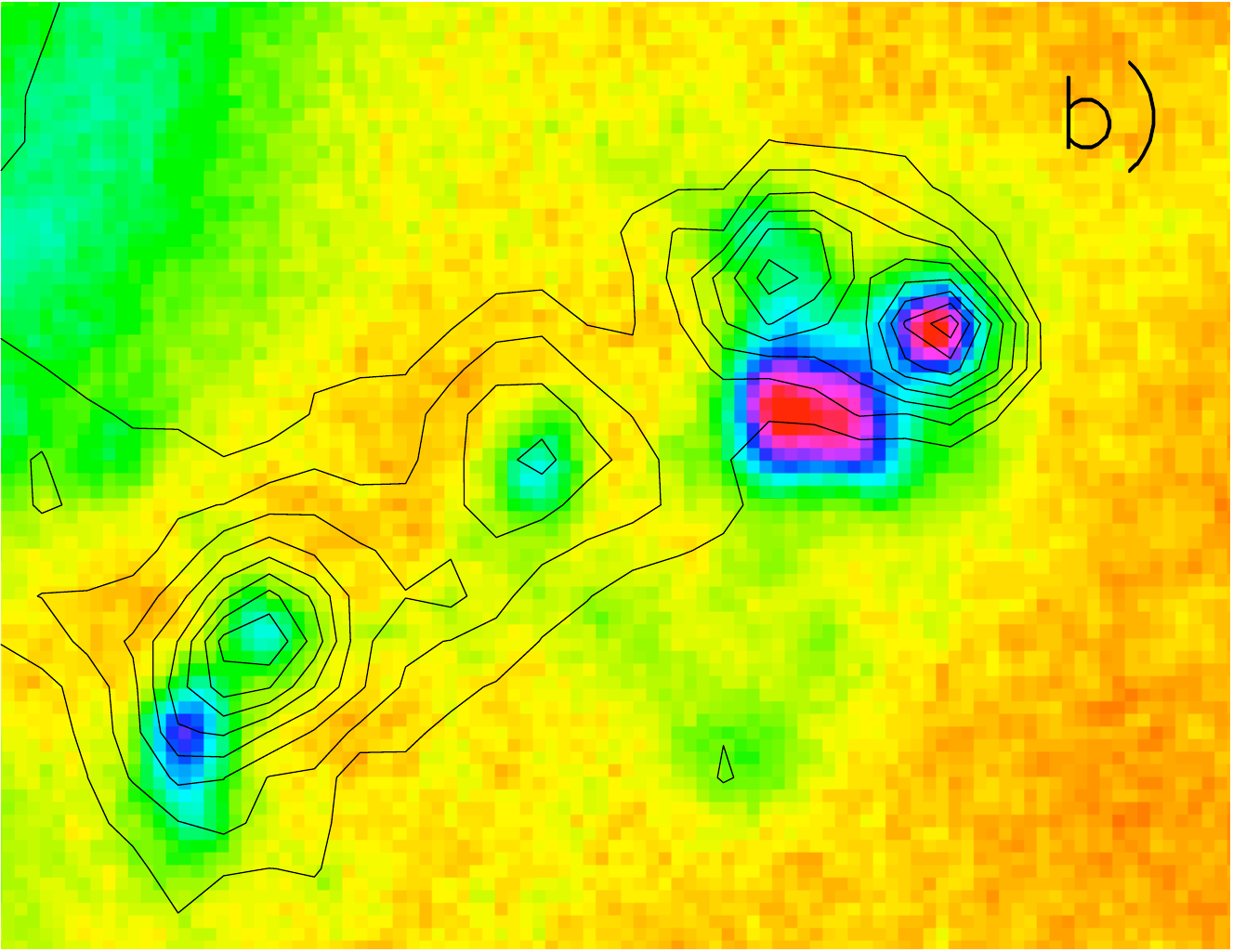}}
\centerline{\plotone{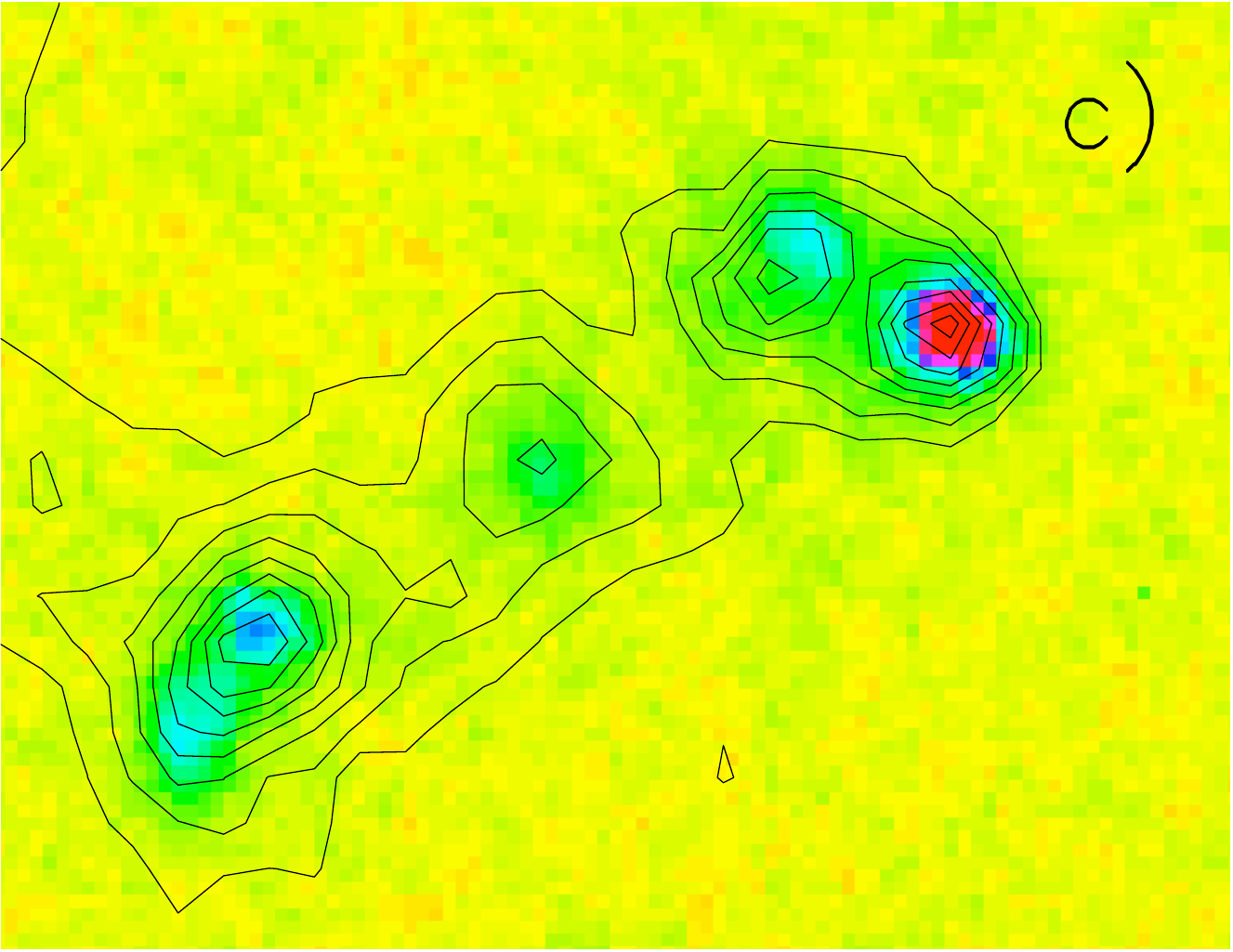}
\plotone{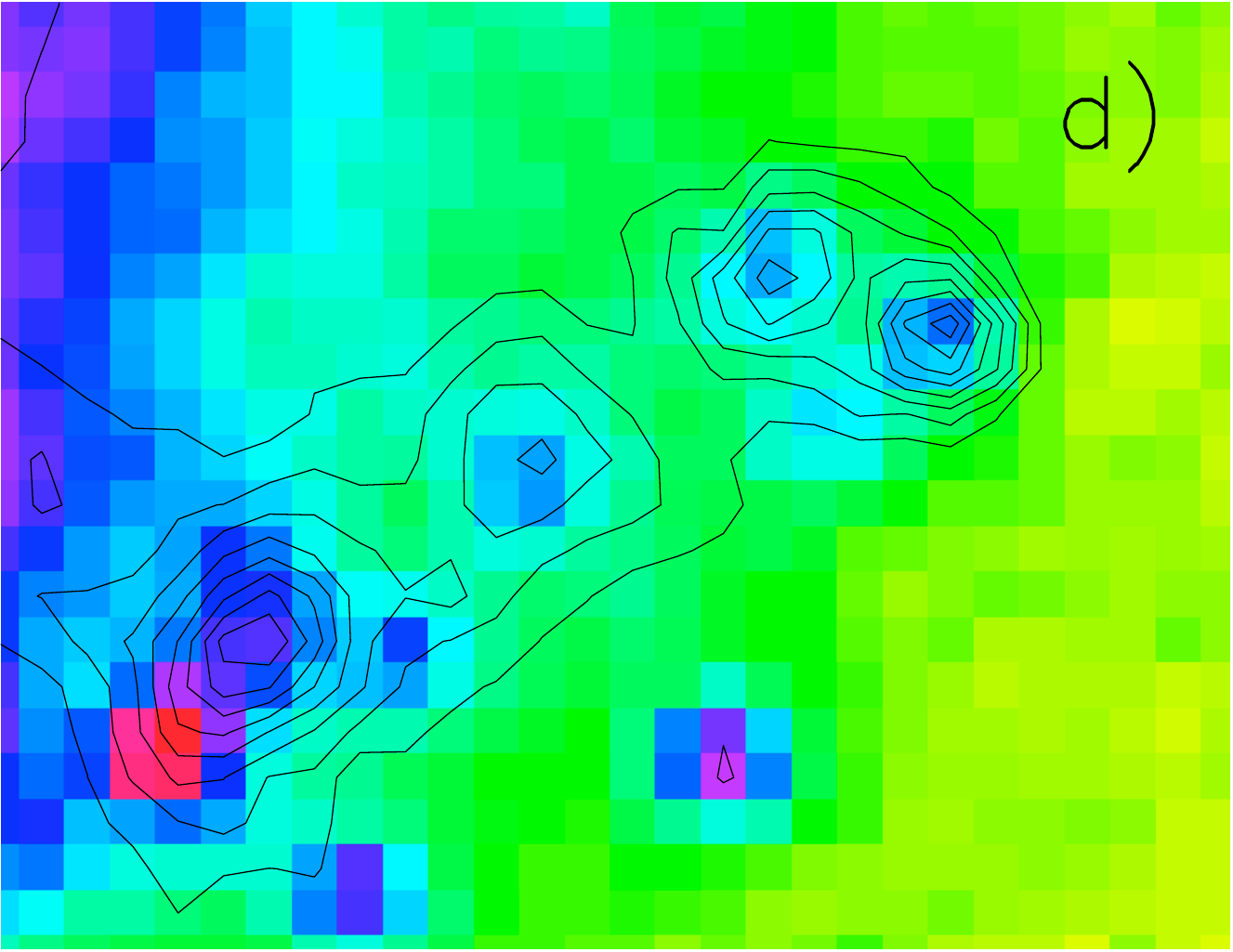}}
\centerline{\plotone{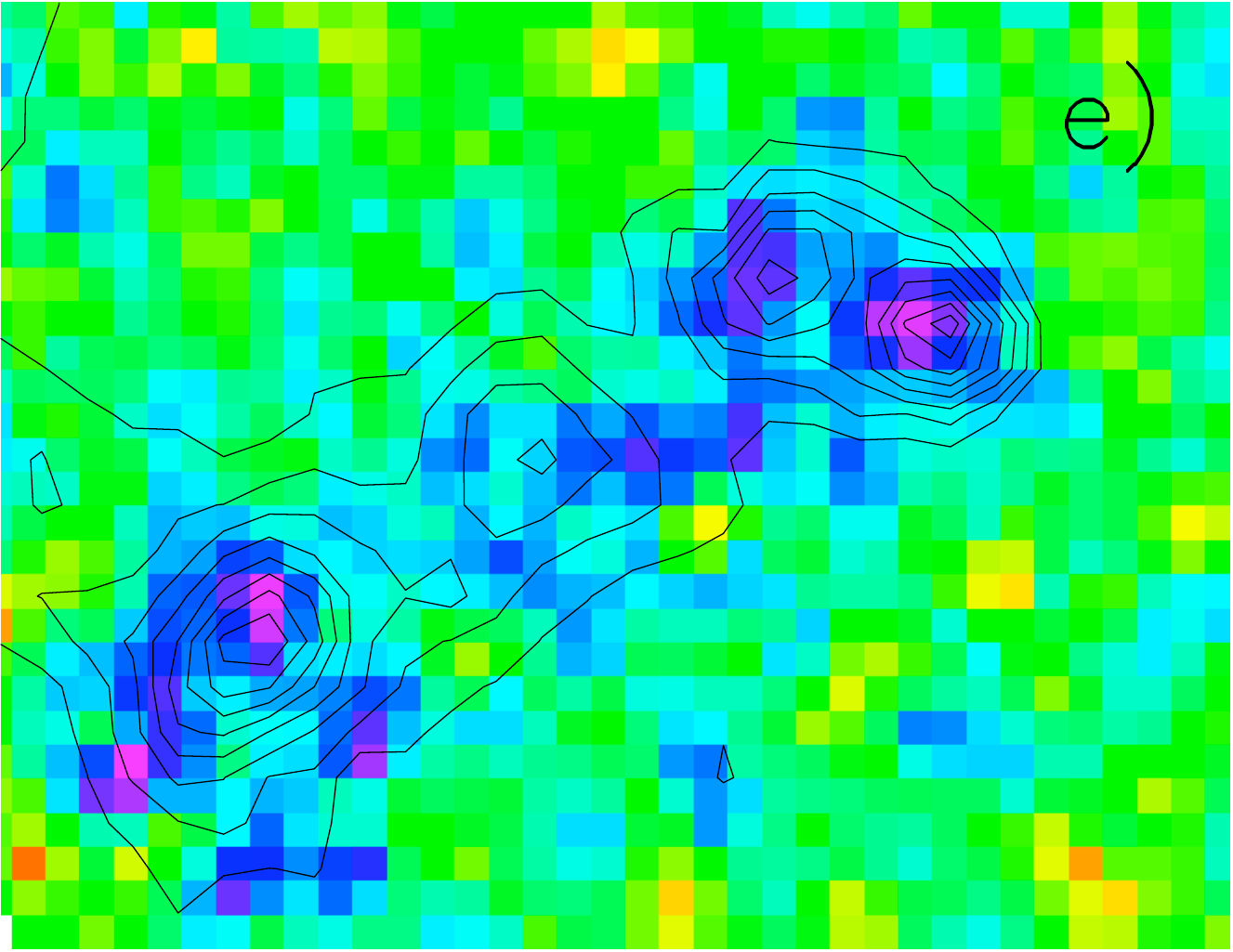}
\plotone{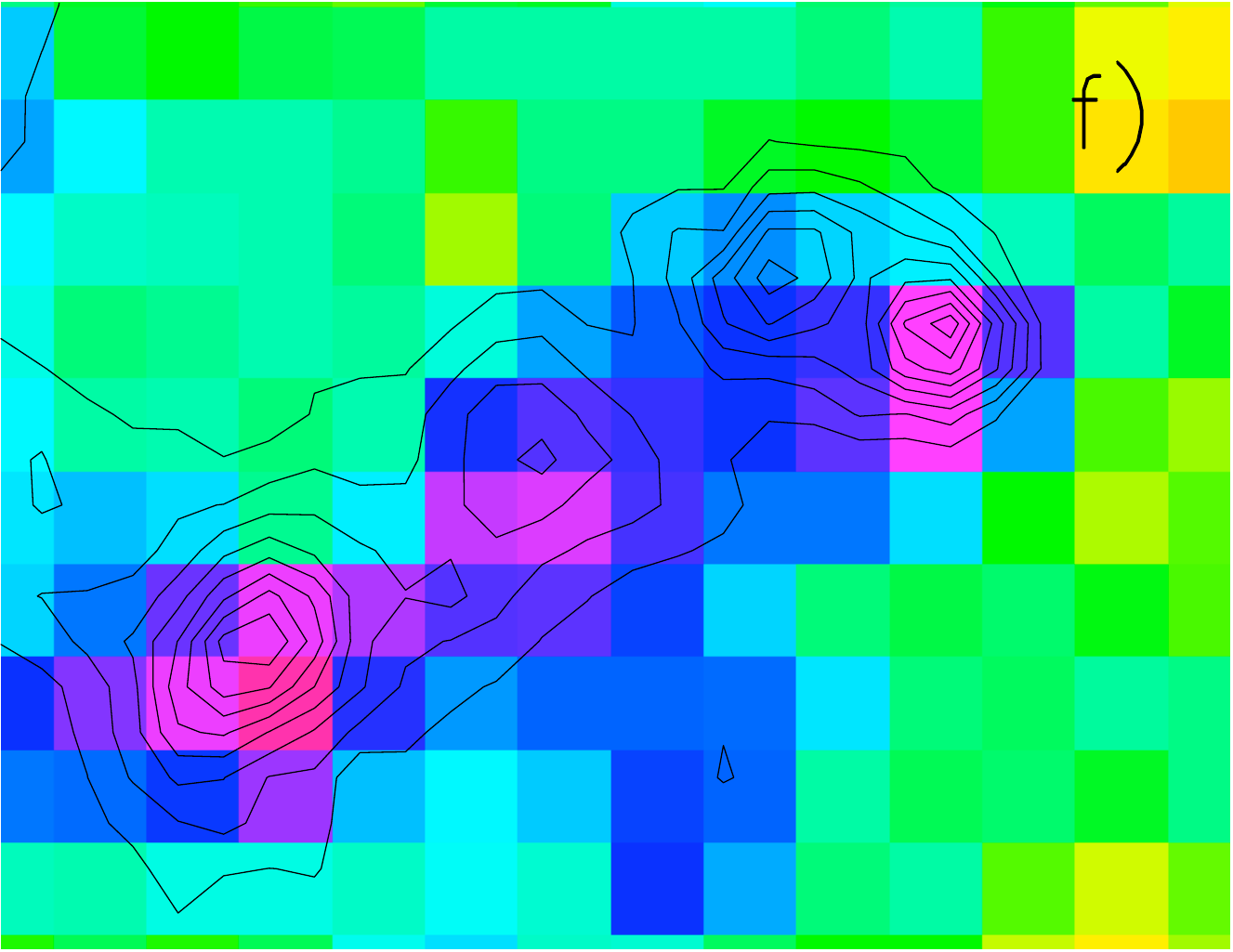}}
\caption{Spitzer observations of J1023+1952: 
a) the B-band image indicating the numbers of the respective knots.
The other panels show the 8$\mu$m contours superimposed on b)  B-band image,
c)  \halpha\ image,
d)  3.5$\mu$m image, 
e)   15$\mu$m image and
f)   24$\mu$m image.
\label{spitzer-irac}}
\end{figure*}

At all observed wavelengths, the only obvious emission within 
J1023+1952 is from the SF knots in the south.  
Figure \ref{spitzer-irac} (a) shows the optical B band image with the
linear arrangement of bright blue knots labelled 1, 2, 3, 4, 5
according to Mundell et al. (2004); we also label additional knots, 1a
and 2a, which were visible in the images of Mundell et al. (2004) but
not specifically identified by them. In 
Fig. ~\ref{spitzer-irac} (b-f)  we show comparisons between  {\it Spitzer} emission at 8$\mu$m
(in contours), the blue, \halpha\
and {\it Spitzer} emissions at 3.6, 15 and 24$\mu$m. 
In general, a good structural agreement between the emissions at all wavelengths
can be seen (exceptions will be discussed below), however  with a 
different relative strengths.
In Table \ref{IRphot} we list the integrated fluxes of the
individual knots with detected emission.

For most knots, there is a very good agreement between the emission at
8 $\mu$m, \halpha\ and the blue.
A striking aspect is the 
absence of 8$\mu$m (PAH + dust continuum) emission from the brightest
optical knot 2 and its strong decrease or absence at knot 5.  
The dust emission seems to be  also missing at longer wavelengths,
albeit, traced with a poorer spatial resolution.
The lack of  8$\mu$m emission is accompanied by a lack of  H$\alpha$
in knot 2, but not in knot 5, which shows strong H$\alpha$ emission.

The 3.6$\mu$m emission, usually associated with the photospheres of
older stars, is visible from all knots, with the strongest emission 
coming from knots 4, 5 and 2a.  
The detected emission is associated only with the knots and no 
additional smoothly distributed emission is detected, thus ruling out the 
presence of an extended underlying old stellar population.
This confirms the previous non-detection of an extended component at near-IR wavelengths
 by Mundell et al. (2004).

The emissions at 15 and 24$\mu$m are also clearly detected, however
with a poorer resolution and S/N ratio.

\subsubsection{The optical/MIR SEDs of the star-forming knots} 

We carried out aperture photometry of the SF knots on images at 
optical wavelengths (B, V, I bands),  \halpha\ (data from Mundell et al. 2004),
and the {\it Spitzer} IRAC bands and 15 \mi\ emission.
Due to the poorer spatial resolution, we could not derive 
the  24$\mu$m fluxes for the individual knots. 
For the aperture photometry, we smoothed the optical images to the resolution
of the 8$\mu$m image (2\arcsec) and carried out
 photometry on this set of images.
We used a circular aperture with a radius of 3.6\arcsec\
centered on the
individual knots indicated in Fig.~\ref{spitzer-irac}.
Because of their small distance, knots 4 and 5 could just barely
be distinguished individually.
The integrated fluxes derived in the aperture photometry are given
in Table~\ref{IRphot}.

\begin{figure}[h!]
\epsscale{0.5}
\plotone{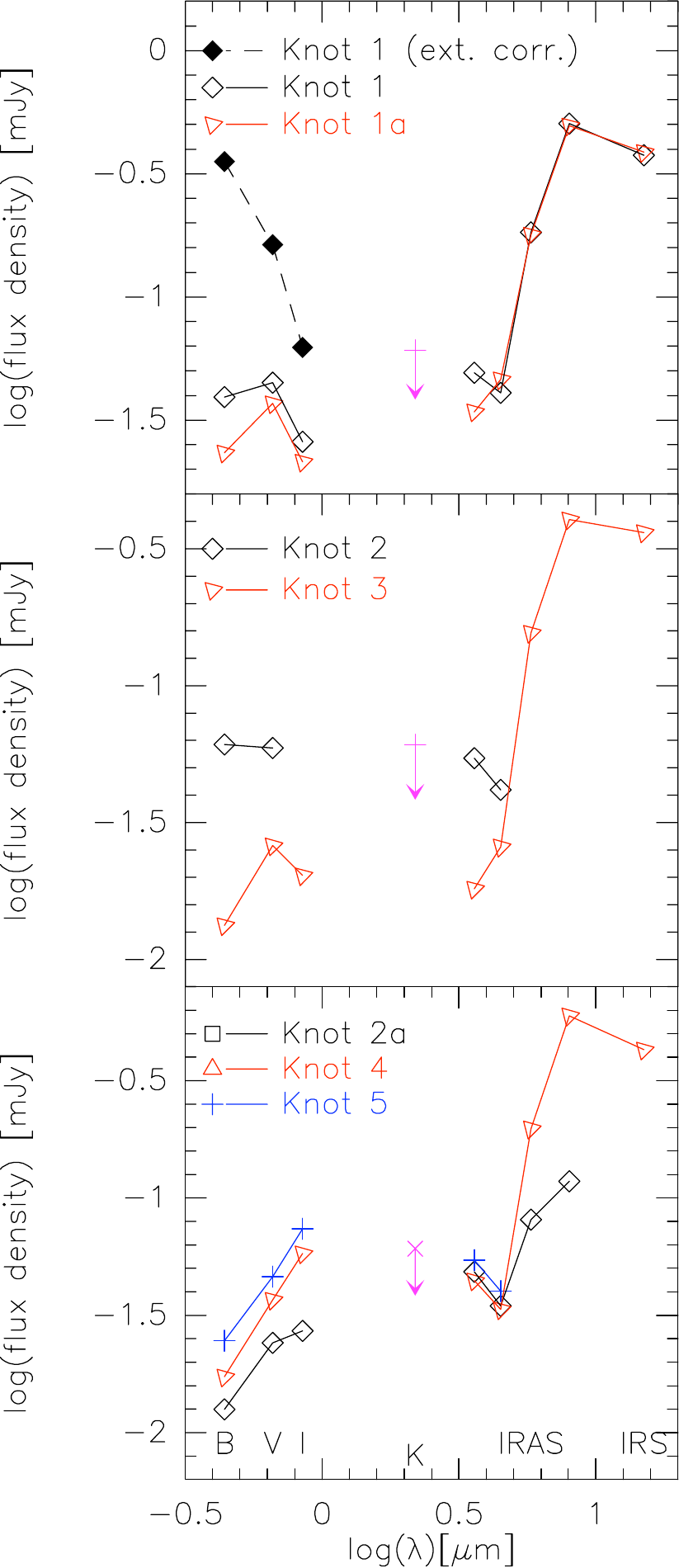}
\caption{
The optical+IR SEDs of the SF knots.
The fluxes used here are listed in Table~\ref{IRphot} and are
uncorrected for extinction, except  for  knot 1, where
we show both the uncorrected (open symbol and full line) 
and the extinction-corrected (filled symbol
and dashed line)
fluxes, using $A_{H\alpha} = 1.4$~mag as measured and
adopting  the extinction curve of Draine (2003) to derive the
extinction at other wavelengths.
The upper limit in the K-band is from
Mundell et al. (2004) and is valid for all knots.
\label{sed}}
\end{figure}

We show in Fig. \ref{sed}  the optical-IR SEDs for the 
individual knots.  For knot 1 we measured the extinction (A$_{\rm H\alpha}=1.4$ mag, 
corresponding to A$_B =2.4$ mag), so that  we can show both the
extinction corrected and uncorrected SED for this knot. 
The knots show clear differences in their SEDs, most likely due
to differences in their SF history and presence of dust.
In the following, we discuss them individually.
We compared the data to  simulations with 
Starburst99 (Leitherer et al. 1999), using both continous and instantaneous
star formation,  and assuming solar metallicity and a Salpeter Initial Mass Function,
in order to set constraints on the age and extinction of the star forming knots.

\begin{itemize}
\item Knot 1 gives a consistent picture of
a young star forming region: the extinction-corrected optical SED shows a very blue color
and \halpha\ and dust emission is present.
The  simulations with Starburst99 yielded ages between
1 and 2 Myr (both for a continous and instantaneous burst) for this 
knot.

\item The uncorrected SEDs of Knots 1a and 3 are very similar to that of knot 1 and 
they also present 
 \halpha\ and dust emission.
Thus, if the  dust extinction for these knots is similar to that in knot 1,  
which is plausible given the small overall variation of the \hi\ column density 
across the SF region (Mundell et al. 2004),
they are also representing regions of very young SF with ages between 1 and 2 Myr.  
The shape of the SED of these knots allows us to constrain the minimum extinction:
The peak in the V-band of the unextincted SED cannot be produced by any synthetic SEDs 
of Starburst99.  We need a minimum extinction of A$_B$ = 1.9 mag in order 
to achieve reasonable agreement between the data and simulations and
derive for this case a starburst age of $4-5$ Myr (both for an instantaneous and a continous burst).
Higher values for the extinction than the reference value of A$_B$ =2.4 mag measured
in knot 1 would indicate a younger age than 1-2 Myr.
Thus, in spite of the uncertainty in the extinction, we can constrain the ages of these 
two SF knots to be at most 5 Myr.

\item  Knot 2 is the brightest knot in the  B-band image and has an 
even bluer color than knots 1, 1a and 3, suggesting a 
similarly young age.
However, knot 2 shows no H$\alpha$, 
indicating that no very young SF is taking place. 
A possible explanation is the absence of dust locally at knot 2.
This hypothesis is consistent with  the absence of long-wavelength emission
from small grains or PAH molecules.
The comparison of Starburst99 simulations to the extinction-free SED
yielded  ages  of 20 to 100 Myr (for instantenous SF) and 50 to 600 Myr (for
a  continous burst).
These relatively old ages would explain the lack of \halpha\ emission.

\item The other knots (knot 4, 5 and 2a) have redder optical SEDs, either due to a higher
extinction or due to an older stellar population.
Knot 4 and 5 show \halpha\ emission, whereas from knot 2a no \halpha\
emission could be detected.
If the extincion at these knots had values of $A_B= 3.5 - 4$ mag the intrinsic
SED would be similar to knot 1 and we would derive a similarly young SF age. 
Alternatively, assuming that the extinction has
the same value as in knot 1, we derive for knots 4 and 5 ages of
10 to 100  Myr  (continous SF) and 5 to 6 Myr (instantaneous burst).
For knot 2a we derive an age of about 5 Myr  for  both burst scenarios.
If the extinction were lower, the ages of the knots would be higher.
In the limiting case of no extinction we derive satisfactory fits for all three
knots for a starburst age of 100 Myr (for an instantaneous burst; for a continous
burst no good fit could be achieved). This value is certainly an overestimate,
at least for knots 4 and 5, because in this case we would not expect any \halpha\
emission.

Interestingly, there are differences in the H$\alpha$ and dust emission:
Whereas knot 4 has clear H$\alpha$ and dust emission,
the dust emission (mostly visible in the 8$\mu$m map,
since the lower spatial resolution of the other wavelengths makes
a clear spatial association difficult) at knot 5 is, if present, at most weak compared to its
H$\alpha$. Such drastic variations between the 8$\mu$m and \halpha\ emission
have been observed occasionally in other intergalactic SF regions as well (e.g. Boquien et al.
2007) and are most likely due to the fact that PAHs can be easily 
destroyed in a strong UV field.

\end{itemize}
 
In summary, we find that the SEDs of all 
 knots, except for knot 2, are consistent with recent SF ($< 10$ Myr).  
Furthermore, assuming that the extinction does not vary substantially over the
SF region, and in particular that nowhere the values are much higher than measured
in knot 1, we find tentative evidence for an age gradient across the knots: the oldest knots, 
with ages most likely between 5-20 Myrs, lie in the south to south-east 
(4, 5, 2a), while the remaining knots (1, 1a, 3)  are very young, at most 2 Myr.
Knot 2 represents an exception in this sequences with most likely an
older age.

\subsubsection{Comparing tracers of star formation}

The use of the 8 and 24\mi\ emission as a tracer for obscured SF has
been studied in numerous galaxies. The 24\mi\ emission
shows a tight relation with the emission
of the ionized gas (Calzetti et al. 2005, P\'erez-Gonz\'alez et al. 2006,
Calzetti et al. 2007) which is independent of metallicity if only the 
\hii regions are taken into account
(Rela\~no et al.
2007). There exists also a relation between the 8\mi\ emission and the
ionized gas emission, but it is less tight and
more dependent on metallicity (Calzetti et al. 2007).

\begin{figure}[h!]
\epsscale{0.5}
\plotone{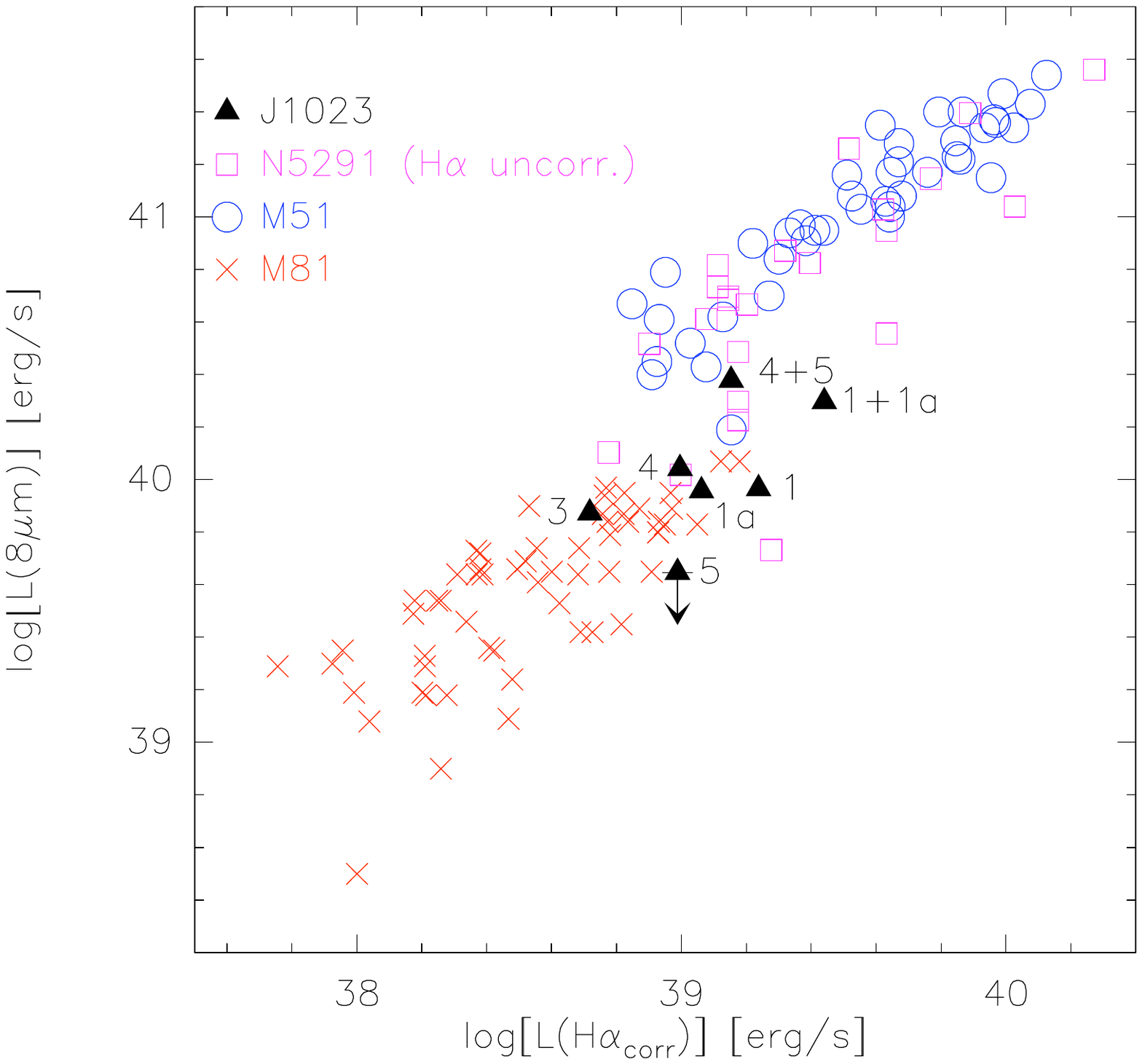}
\plotone{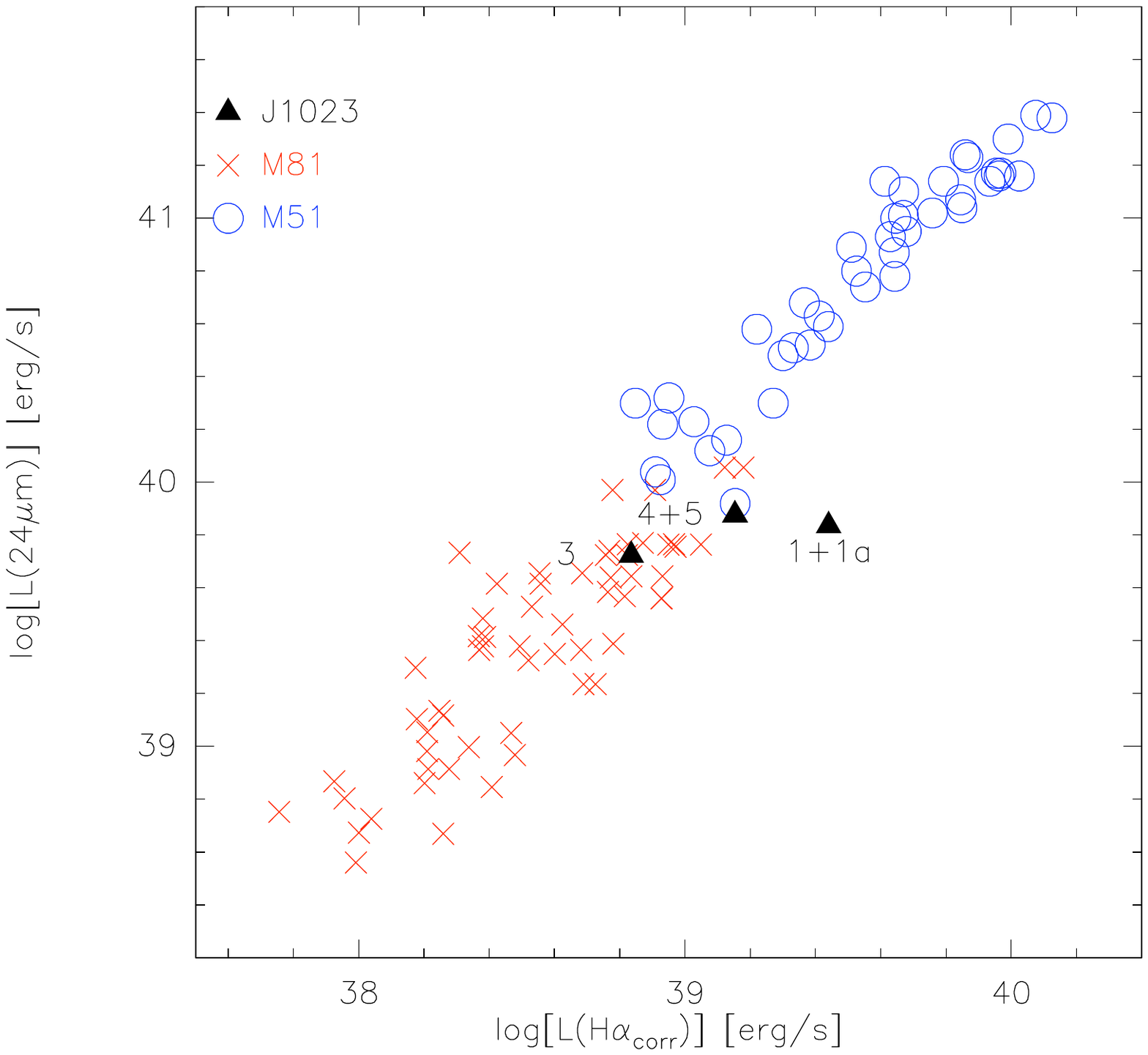}
\caption{{\it Upper panel:} The 8$\mu$m  luminosity 
and the extinction-corrected \halpha \ luminosity (assuming
an extinction of $A_{H\alpha} = 1.4$~mag)
of the knots 1, 1a,
3, 4 and 5, as well as the combined emission of knots 1+1a and
4+5 together with
data of \hii\ regions 
in M51 (Calzetti et al. 2005), M81 (P\'erez-Gonzalez et al. 2006)
and extragalactic \hii\ regions in NGC 5291 (Boquien et al. 2007).
For the latter the \halpha\ luminosity is uncorrected for extinction.
{\it Lower panel:} The 24$\mu$m  luminosity 
and the extinction-corrected \halpha \ luminosity of the knots 1+1a 
(combined), knot 3 and knots 4+5 (combined) together with data
 of \hii\ regions 
in M51 (Calzetti et al. 2005) and M81 (P\'erez-Gonzalez et al. 2006).
\label{spitzer-8-24}}
\end{figure}

Figure \ref{spitzer-8-24}  shows a comparison between the luminosities of the extinction
corrected \halpha, 8$\mu$m and 24 $\mu$m, respectively, 
 for the knots where both
emissions have been detected, together with a comparison
to data from the literature for M81
(P\'erez-Gonz\'alez et al. 2006),
M51 (Calzetti et al. 2005) and, in the case of the 8$\mu$m emission, with the
extragalactic \hii regions in the system NGC 5291 (Boquien et al. 2007). 
The relation between the 8\mi\ and \halpha\ emission of the four knots (1, 1a, 3 and 4) 
in J1023+1952 match those of the \hii regions in M81.
We also include the combined emission of knots 1+1a and knots 4+5 which
were used in the analysis of the 24\mi. 
Knots 1+1a show a slight 
excess of \halpha\, emission, or a lack of 8\mi\ emission, with 
respect to the data points of M51 and M81.

The 24\mi\ emission of knots 1 and 1a as well as of knots 4 and 5
could not be measured individually, so that the combined emission of
both knots is considered whenever a comparison to the 24\mi\ emission is made.
The corresponding values are listed in Table~\ref{IRphot}. Whereas knot 4+5 and knot 3 follow very well the correlation between
the  24\mi\ and the extinction corrected \halpha\ luminosity of the \hii\ regions
in M81 and M51, the emission of knot 1+1a shows an excess of \halpha\
emission, or a lack of of  24\mi\ emission, with respect to the correlation.
The reason for this difference is unclear. 
The uncertainty in the 24\mi\  correction (for the AGN contamination in the center
of NGC~3227)
could  explain part of this discrepancy.
Alternatively, given that the excess of 
\halpha\ luminosity is relatively small (a factor of $\sim  1.5-2$),
a slight  overestimate of the extinction could also 
explain the apparent excess. 

\section{Discussion}

\subsection{Triggering the star formation}

SF in J1023+1952 is restricted to a small region in the southern part. 
Our observations show that the absence of SF in the rest of the cloud is {\it not}  due
to a lack of molecular gas because abundant CO has been
found over the entire object. 
Fig.~\ref{mom0} shows that 
the surface density of the molecular gas 
is relatively uniform over the cloud, 
with values ranging between 3 and 6$\times 10^{20}$ cm$^{-2}$ (Table~\ref{surface_density}).
The column density of the total neutral (atomic and molecular) gas is
rather homogeneous as well with
values ranging between  $\sim$ 1 and 3$\times 10^{21}$ cm$^{-2}$
(see Table~\ref{surface_density}). 
For these values of gas column density,
SF activity has routinely been found in TDGs \citep{braine01}.
The atomic gas surface 
density peaks  in the SF region (see Fig.~\ref{mom0}) but
other places along the \hi\ ridge have similarly high values, so
that this high value  alone cannot explain why
SF is occuring in just one restricted area.

A noticable difference in the gas properties is the
narrower line width both of the CO and the \hi\ 
in the region where SF takes place in comparison
to the rest of the cloud.
Thus, SF is only present where
the gas is dynamically cold, whereas in the rest of the
cloud the higher velocity dispersion of the molecular gas clouds seems to
suppress SF.
%
Our observations thus show that overall gas-richness is not a sufficient condition for
SF, and that the dynamics of the gas have to be considered as well.
Studies by other authors  agree with this result.
In the dwarf galaxy VCC 2062, most likely an old TDG,
SF has also been found
only in the dynamically cold gas (Duc et al. 2007). 
In a study of SF in tidal tails,
Maybhate et al. (2007) found that in order for SF to take place in
tidal arms, a high gas surface density ($\log(N_H) > 20.6$ cm$^{-2}$)
is a necessary but not sufficient condition. In fact, they found several
places in the tidal arms of NGC 4038/39 and NGC 3921 where no 
stellar clusters were found in spite of a high gas surface density.
From our study, we would predict that the gas in the locations without SF has a considerably
higher velocity width than in the places where SF takes place.

The process of SF can empirically be well described by  a 
Schmidt-law for the SFR and a threshold below which SF is inhibited
(Kennicutt 1998a). The SFR in the southern region follows well the
law found by Kennicutt (1998b) for a large sample of galaxies.
Both the gas column density and the SFR per area in this region lie at the lower
end of the range of values found in galaxies (Kennicutt 1998b) and thus
close to the values where
thresholds become important.


There are two kinds of thresholds in gas column density  that have to be considered
(Elmegreen 2002).
A minimal gas column  density (of about 6 \msol/pc$^2$ for a typical radiation
field and a solar-like metallicity, Elmegreen 2002) is required to provide a high enough
pressure to support a cool phase of \hi\ clouds, necessary  as a first
step for SF. Apart from this, a critical column density exists below which
large clouds cannot form, either due to destructive Coriolis forces (the Toomre
criterion, Toomre 1964) or large-scale shears (Elmegreen 1987, 1991, Hunter, Elmegreen \&
Baker 1998).

In J1023+1952 the gas column density ranges between 10 and
20 \msol/pc$^{-2}$, which is above typical values for the minimum gas column densities.
Further and strong evidence showing that the minimum gas column density
is exceeded in the entire cloud 
comes from the fact that the  gas is sufficiently dense  to support a high 
molecular gas fraction.  
We do not find evidence  that a threshold based on a 
critical density can explain the differences in SF in the north and the
south of the cloud. The Toomre criterion predicts a critical
surface density, $\mu_{\rm crit} = \frac{\sigma \kappa}{\pi G}$,
where $\sigma$ is the velocity dispersion within the clouds,
$\kappa$ is the epicyclic frequency
and $G$ is the gravitational constant. If we assume that the 
velocity gradient found in the gas is due to rotation, we derive little variation
for $\mu_{\rm crit}$, with values between 4 and 5 \msol/pc$^2$, within the cloud, and a trend
for the lower values being situated
in the northern part, so that this criterion cannot explain the absence
of SF in the north. A similar conclusion results from the  stability criterion
based on the shear which yields a critical column density
of $\mu_{\rm crit} = \frac{2.5 A \sigma}{\pi G}$ (Elemegreen 1993, Hunter, Elmegreen \&
Baker 1998), with A being the Oort constant A.
The critical gas column densities in this case are 
slightly higher in  the northern part, however all values are
below 1 \msol/pc$^2$ so that they cannot explain  the lack of SF.

Thus, there must be a different mechanism that inhibits SF in the northern part, e.g. large
scale turbulence which could prevent SF if the motions continously force the gas to break up to pieces that are smaller than a thermal Jeans mass (Padoan 1995, Elmegreen 2002).
Large-scale turbulence would be consistent with the higher velocity width in this
region. Alternatively, SF could have been triggered in the
cool molecular gas present in the cloud, but only in the southern part, possibly due
to gas compression related to the tidal interaction.

\subsection{The nature and origin of J1023+1952}

J1023+1952 is an intriguing object whose origin is a matter of debate. 
The  main scenarios are that J1023+1952 is (a) 
a classical dwarf galaxy, either involved in the interaction
or seen as a chance projection,
(b)  a small spiral galaxy being accreted and tidally disrupted by the NGC3226/7 system,
(c) made of gas from the disk of NGC 3227 expelled by the impact of a third body,
or (d) a potential TDG formed from tidal 
debris extracted from the gaseous disk of NGC~3227. We discuss these possiblilies in turn.

\subsubsection{A preexisting dwarf galaxy?}

One key property that distinguishes ``classical'' dwarf galaxies (i.e. 
not formed from recycled gas) from 
TDGs is their metallicity.
Since the latter are made from recycled gas, their metallicities
are close to those of the parent galaxies. Therefore, 
TDGs do not follow the magnitude-luminosity relation found for
classical dwarf galaxies, on the contrary their metallicities lie
in a narrow range of 12+log(O/H) = 8.3-8.6 \citep{duc00}.
The total brightness of  J1023+1952 of $M_B \sim
-15.9$ (corrected for  an extinction of $A_B = 2.4 $~mag derived
from our spectroscopy)
 would predict a metallicity of  12+log(O/H) $\sim$ 8.0.
This is well below the metallicity estimated from the optical spectra
(Sect. 3.2) showing that  J1023+1952 cannot be a ``classical''
dwarf galaxy.
The detection of large quantities of CO further confirms this high
metallicity of the neutral gas.
At low metallicities of 12+log(O/H) $\sim$ 8 or below
the detection rate of classical dwarves in CO is
very low  \citep[e.g.][]{taylor98},
indicating that CO is no longer a good tracer of the molecular gas content.
In J1023+1952, in contrast, we derived a high molecular-to-atomic gas mass ratio of
60\%, which is even  higher than the range of values found
for TDGs \citep{braine01,lisenfeld02} of
$\sim10-50\%$.
Thus, both observations, the high metallicity and the high molecular gas abundance,
lead us to discard the hypothesis that  J1023+1952 is a preexisting dwarf galaxy.

\subsubsection{An infalling, disrupted small spiral galaxy?}

Deep optical images of the system (see Fig. \ref{tidal-stream}, or Fig. 3 (left) in
Mundell et al. 2004) show faint loops
of emission around NGC 3227 indicating the presence of stellar stream.
This clearly indicates that something is being accreted onto NGC 3227.
 J1023+1952 is placed at the 
intersection of two ends of such  loops
suggesting that its existence might be causually related to them.
A third object (a small spiral galaxy) falling into the system and being disrupted
by the gravitational field of NGC 3227/6 could be the origin of the
stellar streams.
The stellar content in this accreting object is then being  disrupted into long tidal streams
around the system. The dissipative gaseous component, on the other
hand, does not follow the stellar
distribution exactly, but is more concentrated at the position where the streams cross
so that the gas clouds collide and loose their energy. This could explain
the presence of J1023+1952 at just this position.
The collision of the gas clouds would explain furthermore the onset of
recent star formation due to the compression of the gas.

The infalling galaxy would however have to be
much smaller than NGC 3227 in order to understand that it is
disrupted while  NGC 3227 is only very little disturbed.
Taking, e.g., a factor of 10 as a reasonable mass and luminosity difference,
we would expect an object with a brightness of $M_B \sim -17$ mag.
For such a small object, the expected metallicity is much
lower than observed, only 12+log(O/H) $\sim $ 8.0.  Turning the argument
around, we predict,
using the luminosity-metallicity relation, from the
metallicity of J1023+1952 a luminosity of  $M_B \sim -19$ mag for the infalling
object, which
is practically the same as for NGC 3227. Such a large object, if
in interaction with the system, would be expected on 
cause major
damage to NGC 3227 as well.
Therefore, J1023+1952 cannot be made of material from a third infalling
object, it needs to be made of gas from NGC 3227 or NGC 3226.

\subsubsection{Collisionally expelled gas from NGC 3227?}

A further alternative is that a small  galaxy has passed
through the disk of NGC 3227 and has expelled part of the gas from the disk 
via cloud-cloud collisions, in the same way as gaseous bridges are formed
between galaxies envolved in head-on collisions.
In this case J1023+1952 would be made of 
gas from the disk of NGC 3227 which would explain its high metallicity.
The position-velocity diagram along the major axis of NGC 3227
shown in Fig. ~7 in Mundell et al. (1995) shows that the gas in J1023+1952 
corresponds roughly to the atomic gas  that seems be missing from the
central region of NGC 3227, which is consistent with this scenario.
However, in a direct collision  one would expect the formation of a gaseous bridge
between the two objects, as e.g. observed in the Taffy
galaxies (Condon et al. 1993, Braine et al. 2003, 2004),  and not only emission at two discrete
velocities, as observed in NGC 3227 and J1023+1952.
Therefore, although at first sight an attractive possibility, J1023+1952
cannot be made of collisionally expelled gas from the disk
of NGC 3227.

\subsubsection{A potential TDG?}

Alternatively, J1023+1952 could be a TDG, an object formed from tidal debris stripped
from  the disk of NGC 3227 during the interaction, explaining in this way
naturally the high metallicity and molecular gas content.
In comparison to other TDGs which normally lie at the end of tidal tails
(e.g. Duc et al. 2004), 
the position of J1023+1952 at the base of the northern plume is however unusual.
The young age of its stellar populations  suggests that
it has formed recently and the high metallicity and high molecular 
gas content imply that it has
formed from material coming from the main disk. Possibly it has formed
in a second phase of interaction, after a first one which expelled
the outer \hi\ disk and formed the \hi\ tidal tails.
This late and recent formation might explain its unusual 
position relatively close to  NGC~3227.
Evidence for previous removal of atomic gas  from the disk of NGC~3227
comes from the relatively strong CO emission from the south-eastern part of the outer disk
of NGC~3227  (see Fig.~\ref{spectra}). The  high ratio of molecular-to-atomic gas surface density
(2N(H$_2$)/N(HI) = 1.2) at this position is unusual for outer disks and could
be an indication of  removal of atomic gas in a previous phase of the
interaction.

\begin{figure}[h!]
\epsscale{1.0}
\plotone{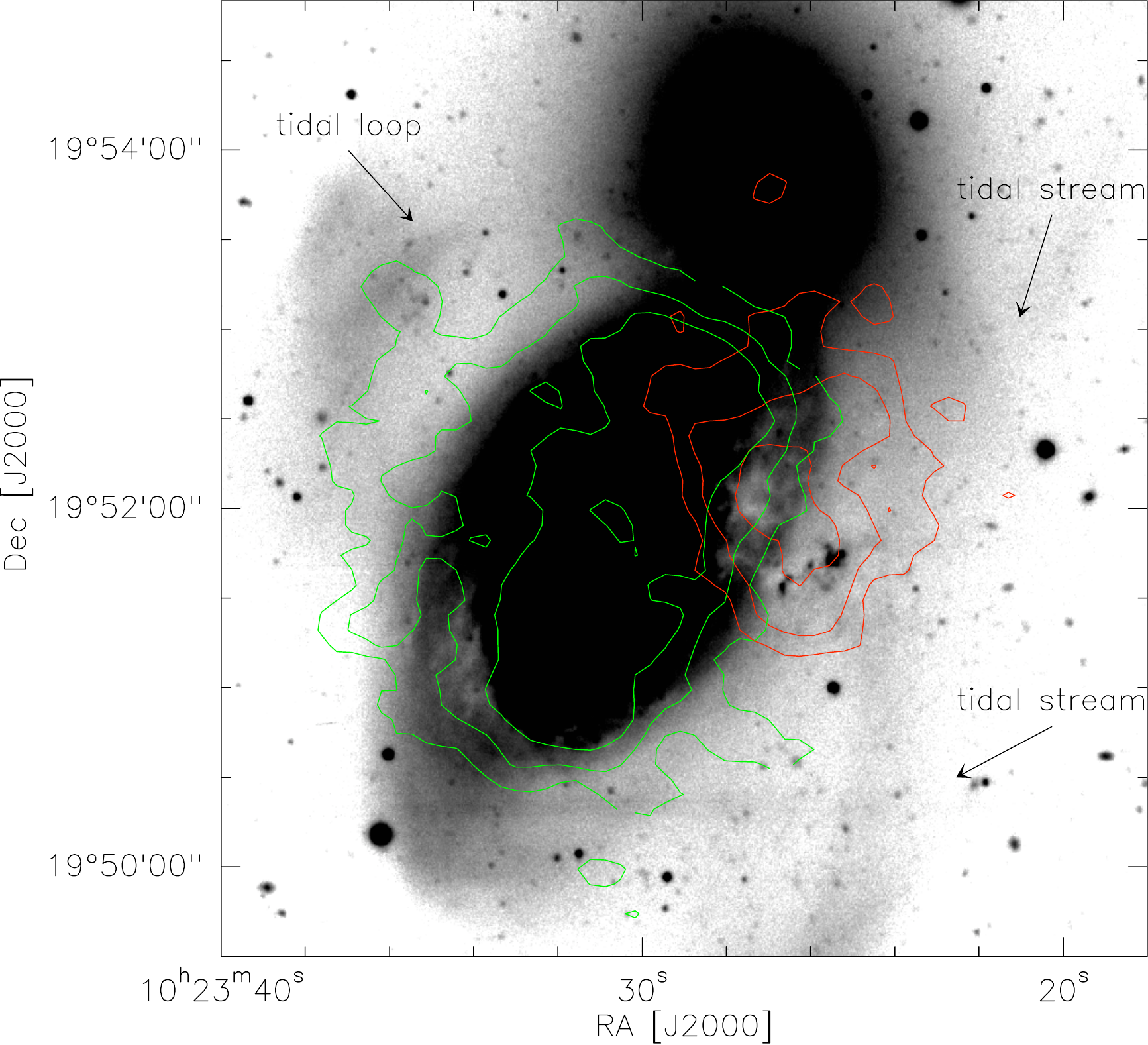}
\caption{A deep B-band image of Arp 94 showing the tidal streams
around the object.  J1023+1952 is situated at the intersections
of two ends of a stream, close (in projection) to the outer disk
of NGC 3227. Furthermore, the extinction caused towards the 
disk by the dust
associated with the neutral gas in J1023+1952 is clearly seen.
The green contours show the HI emission associated with the
disk of NGC~3227 and the red contours those associated with
J1023+1952 (both from VLA C-array, Mundell et al. 1995).
\label{tidal-stream}}
\end{figure}

A suggestive scenario is that the tidal streams, described in the  Sect. 4.2.2,
are debris from a previous phase of the interaction. They
could either have their origin in NGC 3227, representing 
material from the disk that was expelled in an earlier stage of the interaction,
or, in a more speculative but not unrealistic scenario, be related to the formation
of NGC~3226. 
This dwarf
elliptical has a lot of unusual features (Appleton et al. in prep.)
including a highly peculiar box-like outer light distribution,
a rapidly rotating nuclear core, and most
importantly, long stellar streams emanating from it on deep optical images
(see e.g. Fig.~\ref{tidal-stream}). All these factors suggest that
either NGC~3226 is itself a merger product, or at the very least,
it has accreted material from NGC~3227.
J1023+1952 is situated at the crossing of two tidal streams.
Although it is possible that this spatial coincidence  is due
to a projection effect an alternative scenario is that  
J1023+1952 has been formed {\it due to} the crossing of two tidal
streams. In this case, the recycled nature of the gas
presumably associated with the tidal streams
would explain the high metallicity and high molecular gas content.
Our VLA observations  (see Fig. ~\ref{tidal-stream}) 
indeed show some indications of gas associated with the tidal streams, especially 
at the eastern side, supporting this possibility. However, deeper VLA
observations are required
in order to find out  whether the stellar streams really contain gas  as
required in this scenario.


\section{Conclusions and summary}

We presented a comprehensive study of the ISM of J1023+1952, based on new 
millimeter  and 
infrared data, with the goal of determining its nature and origin.
The main results and conclusions are:

1) \co\ and \CO\ emission was detected for the first time and
shown to be extended across the entire 
\hi-emitting  cloud. The intensity-weighted  velocities and the line widths of
the CO and \hi\ emission agree well.  

2) The optical spectroscopy provided measures of the extinction
and metallicity of the brightest SF knot (knot 1). We
derived an extinction of 1.4 mag for the \halpha\ line (corresponding
to $A_B =2.4$) and a metallicity of
12+log(O/H) = 8.6 $\pm$ 0.2 showing that the gas is pre-enriched.

3) We identified seven SF knots in optical and {\it Spitzer} images,
and presented the results of aperture photometry. The SEDs of these
objects were indicative of young star formation (ages between 1 and
10-20 Myr), although they
showed surprising differences. We identified a tentative age gradient
from the south-east, where the two knots (knot 4 and 5) had the reddest color, 
to the north-west (knots 1 and 1a), where a 
very young stellar population was present.
One  knot (knot 5) had \halpha\ emission with no associated
8\mi\ emission. Knot 2 has  the strongest blue emission but
showed no \halpha\  and no dust emission in spite of its blue color.
A possible reason might be the lack of dust which would lead to an overestimate
of the extinction and thus to a too blue color.

4) We studied the properties of the 8\mi\ and 24\mi\ emission in
comparison with the extinction corrected \halpha\ emission, in
order to investigate the dust properties and the use of these two
emissions as SF tracers.
Neither the 8-to-\halpha\ ratio nor the 24-to-\halpha\ ratio of most
knots showed a  significant 
difference in comparsion to
\hii\ regions in M81 and M51. Only for the combined emission of
knots 1+1a we found a slight
excess of \halpha\ with respect to the 24\mi\ luminosity.
%

5) We discussed various possible scenarios for the origin of J1023+1952.
A strong constraint is the high metallicity, which is supported by
the abundant CO emission, much higher than 
expected from its luminosity if J1023+1952 were a preexisting dwarf 
galaxy. Based on this fact, we can exclude a chance superposition of 
a dwarf galaxy or an infalling, tidally
disrupted  small galaxy  as the origin of J1023+1952.
Instead, we conclude that J1023+1952 must have   formed from recycled gas from
the parent galaxies. 

\acknowledgments

We would like to thank the referee for useful and detailed comments on the 
draft, as well as Almudena Zurita, Monica Rela\~no,
Beatriz Ruiz, Vicent Mart\'\i nez and Cesar Husillos  for 
carrying out the observations at Calar Alto. We would furthermore
like to acknowledge the program Calar Alto Academy that allowed
these observations to be taken. We are grateful to  H. Teplitz (Caltech) for help with
IRS peak-up imaging used in this paper and to T.  Jarrett (IPAC) for useful discussions.
UL thanks IPAC for its  hospitality during a summer visit where part of this work was
done and acknowledges financial support from the research project
AYA 2005-07516-C02-01 and ESP 2004-06870-C02-02 from the
Spanish Ministerio de Ciencia y Educaci\'on and from the
Junta de Anaduc\'\i a.
CGM acknowledges financial support from the Royal Society and the
RCUK. 

{\it Facilities:} \facility{CLFST ()}, \facility{CAO:2.2m ()}, \facility{Spitzer ()}, \facility{IRAM:30m ()}




\clearpage



\begin{table*}[h!]
\caption{Extinction corrected line intensities of knot 1 with respect 
to H$\beta$ = 100}
\label{optical_lines}
\begin{tabular}{cc}
\hline
\hline
H$\beta$ & 100 \\
H$\alpha$ & 286$\pm$16\\
$[$OII$]_{\lambda 3727}$ & 430$\pm$16 \\
$[$OIII$]_{\lambda 4959}$ & 54$\pm$12 \\
$[$OIII$]_{\lambda 5007}$ & 133$\pm$4 \\
$[$NII$]_{\lambda 6584}$ & 85$\pm$2 \\
$[$NII$]_{\lambda 6448}$ & 37$\pm$3 \\
$[$SII$]_{\lambda 6731}$ & 43$\pm$6 \\
$[$SII$]_{\lambda 6716}$ & 34$\pm$3 \\
\hline
\end{tabular}
\end{table*}

\begin{table*}[h!]
\caption{Aperture integrated optical and IR flux densities of the star forming 
knots} 
\label{IRphot}
\footnotesize
\begin{tabular}{lccccccccccc}
\hline
Knot &  r$_{\rm aperture}$ & S$_{B }$ & S$_{R}$ & S$_{I}$ & S$_{3.6}$ & S$_{4.5}$ &  S$_{5.8}$ & S$_{8.0}$ &S$_{15}$ & S$_{24}$   & S$_{H\alpha}$  \\
 & [\arcsec] & [$\mu$Jy] &[$\mu$Jy] &[$\mu$Jy] &[$\mu$Jy] &[$\mu$Jy] &[$\mu$Jy] &[$\mu$Jy] 
&[$\mu$Jy] &[$\mu$Jy] & [$\rm 10^{-15} erg s^{-1} cm^{-2}$] \\
\hline
1  & 3.6  & 39.3  & 44.9  & 25.8 & 49.3 & 40.8 &	183 & 505 &  376 & -- & 9.6\\
1a & 3.6 & 23.3  & 36.9 & 21.4 & 34.2 & 46.2 &	179 & 495 &  387 & -- & 6.4\\
2  & 3.6 & 60.9  & 59.2  &  --      & 54.4 & 41.6 &	 --        & ---    &  --    & -- & -- \\
2a & 3.6 & 12.6  & 24.2 & 27.2 & 48.5 & 34.7 &	 80.9  & 118 &  --    & -- & -- \\
3  & 3.6 & 13.3  & 26.0  & 20.3 & 18.1 & 25.8 & 156 & 404/590\tablenotemark{1} &  363 & 855\tablenotemark{1}& 2.9/3.8\tablenotemark{1}\\
4  & 3.6 & 17.3  & 36.8  & 58.0 & 44.4&33.5 & 197 & 598 &  428 &  --     & 5.5\tablenotemark{2}\\
5  & 3.6 & 24.7  & 46.3  & 73.9 & 54.4 & 40.1 & --     & $<$ 250   &  --   &  --                         & 5.4\tablenotemark{2}\\
1+1a & 7.2&   --  &     -- &   --  &	  --  &	  --  & --                       & 1080     & --   & 1104             & 15.3\\
4+5 & 7.2 &   --   &   --   &   --  &	 --   &	  --  & --                      & 1290   & --   & 1216                & 7.9\tablenotemark{2}\\
\hline
\end{tabular}

Aperture integrated optical and IR flux densities of the star forming 
knots in J1023+1952 (see Fig. \ref{spitzer-irac} for the knot 
identifications). Flux denities are  uncorrected for extinction. 
The \halpha\ fluxes include a correction factor of 0.77, derived from
our spectroscopy (see Table \ref{optical_lines}), for NII emission.
For the {\it Spitzer} fluxes we applied aperture correction, based
on the assumption that the knots are point sources.
The correction factors used were 
(1.13, 1.14, 1.20, 1.24, 1.66) at  (3.6, 4.8, 5.8, 8,15$\mu$m) for the 3.6\arcsec\ aperture,
(1.16, 1.90) at (8, 24$\mu$m) for the 4.9\arcsec\ aperture
and (1.07, 1.60) at (8, 24$\mu$m) for the 7.2\arcsec\ aperture.
The error of the fluxes are estimated to be about 10\% at all wavelengths due to
uncertainties in the background subtraction, and 30\% at 24$\mu$m due 
to the more elaborate image processing that was necessary (see Sect. 2.3).

\tablenotetext{1}{aperture radius of 4.9\arcsec}

\tablenotetext{2}{The sum of the fluxes of knot 4 and 5 are slightly higher that the integrated
flux from both knots because of contamination with the adjacent knot in the 3.6\arcsec\
aperture (see Sect. 3.2.2).}

\end{table*}             


\begin{table*}[h!]
\caption{Surface densities in various regions of J1023+1952}
\label{surface_density}
\begin{tabular}{lcccc}
Region$^1$ & N(H$_2$) & N(\hi)$^2$ &  2N(H$_2$)/N(\hi) & 2N(H$_2$)+N(\hi) \\
 & ($10^{20}$ cm$^{-2}$) & ($10^{20}$ cm$^{-2}$) &  & ($10^{20}$ cm$^{-2}$) \\
\hline
1 & 3    & 12   &  0.5     &  18 \\
2 & 6    & 15   &  0.8     &  27 \\
3 & 5    & 6     &  1.7     &  16 \\
\hline
\end{tabular}

\tablenotetext{1}{The positions of the region are indicated in Fig. \ref{mom0}}

\tablenotetext{2}{The \hi\ surface density was obtained from the \hi\ C-array data from Mundell et al. (1995)
which have a similar spatial resolution (20.47\arcsec $\times$ 18.25\arcsec) as  the
CO data.}

\end{table*}


\clearpage

\end{document}